\documentclass[onecolumn]{aastex631}
\usepackage{textcomp}
\usepackage{graphicx}
\usepackage{amssymb}
\usepackage{amsmath}
\usepackage{epstopdf}
\usepackage{gensymb}
\usepackage{natbib}
\usepackage{appendix}
\usepackage{tabu}
\usepackage{color}
\usepackage{multirow}
\usepackage{booktabs}
\usepackage{ulem}
\usepackage{threeparttable}
\definecolor{darkblue}{rgb}{0.17, 0.49, 0.72}
\definecolor{darkgreen}{rgb}{0.0, 0.4, 0.0}
\newcommand{\sref}[1]{Section~\ref{#1}}
\newcommand{\fref}[1]{Figure~\ref{#1}}
\newcommand{\eqn}[1]{Equation~(\ref{#1})}

\newcommand{\app}[1]{Appendix~\ref{#1}}
\newcommand{\mpsa}{MPS-ATLAS}

\begin{document}
\title{Sensitivity of spectral lines to granulation: The Sun}
\shortauthors{Sowmya et al.}

\correspondingauthor{K.~Sowmya}
\email{sowmya.krishnamurthy@uni-graz.at}

\author[0000-0002-3243-1230]{K.~Sowmya}
\affiliation{Institute of Physics, University of Graz, Universit\"atsplatz 5, 8010 Graz, Austria}

\author[0000-0002-8842-5403]{A.~I.~Shapiro}
\affiliation{Institute of Physics, University of Graz, Universit\"atsplatz 5, 8010 Graz, Austria}
\affiliation{Max-Planck-Institut f\"ur Sonnensystemforschung, Justus-von-Liebig-Weg 3, 37077 G\"ottingen, Germany}

\author[0009-0009-3020-3435]{V.~Vasilyev}
\affiliation{Max-Planck-Institut f\"ur Sonnensystemforschung, Justus-von-Liebig-Weg 3, 37077 G\"ottingen, Germany}

\author[0000-0002-0929-1612]{V.~Witzke}
\affiliation{Institute of Physics, University of Graz, Universit\"atsplatz 5, 8010 Graz, Austria}

\author[0000-0002-8863-7828]{A.~Collier~Cameron}
\affiliation{Centre for Exoplanet Science, SUPA School of Physics and Astronomy,
University of St Andrews, North Haugh, St Andrews KY16 9SS, UK}

\author[0000-0002-3418-8449]{S.~K.~Solanki}
\affiliation{Max-Planck-Institut f\"ur Sonnensystemforschung, Justus-von-Liebig-Weg 3, 37077 G\"ottingen, Germany}

\begin{abstract}
The intrinsic variability of stars, due to acoustic oscillations, surface granulation, and magnetic activity, introduces radial velocity (RV) jitter in spectral lines, obscuring true planetary signals and hindering the detection of Earth-like planets. Granulation is particularly challenging, as it affects even the most inactive stars introducing substantial signals, with amplitudes up to 1\,m\,s$^{-1}$. Disentangling granulation-induced RV jitter from signal caused by planetary reflex motion requires reliable models of stellar granulation. In this study, we present a new approach for calculating sensitivities of spectral lines to granulation. We simulate near-surface convection with 3D radiative MHD code MURaM and calculate high-resolution  emergent spectra with the radiative transfer code MPS-ATLAS. We then show that the spatial and temporal variability of spectral lines due to granulation are almost identical, and introduce a novel methodology that uses spatial variability at a single moment in time to compute their temporal variability. This approach significantly reduces computational costs. We apply our approach to analyze the response of lines from neutral and singly ionized elemental species to solar granulation. We find a clear distinction between the two groups of lines: those from neutral elements tend to show stronger variations in line strength, whereas those from singly ionized elements exhibit larger variations in central wavelength. These results enable the development of spectral line masks tailored to granulation sensitivity, offering a promising strategy to reduce granulation-induced RV noise and improve exoplanet detection.
\end{abstract}

\keywords{Stellar photosphere -- Solar granulation -- Radiative transfer -- Exoplanets -- Radial velocity}

\section{Introduction}
The convective transport of energy in the outer layers of stars gives rise to the surface phenomenon of granulation. It is a pattern of usually bright granules where hot plasma is rising and dark intergranular lanes where the cool plasma is descending. On the solar surface there are about a few million granules at any given time, with individual granules typically having a size of about 1,000\,km and evolving on timescales of 5–10 minutes \citep[see][for a review of solar granulation]{Nordlundetal2009}. While granulation can be directly observed on the Sun since the solar surface can be spatially resolved, we cannot directly resolve it on other Sun-like stars. However, there are other well-known signatures of stellar granulation. The continuous formation and disappearance of granules across different locations on the surface lead to variations in the disk-integrated brightness. Brightness variations due to granulation on timescales below a day are observed and explained for the Sun \citep[see e.g.][]{Seleznyovetal2011,Solankietal2013,Shapiroetal2017}, and are also measured on other stars where they are referred to as photometric flicker \citep{Bastienetal2014}. 

Granulation also affects the positions and profiles of spectral lines. Hot and bright rising granules contribute more photons to the observed spectrum relative to the cool and dark sinking intergranular lanes, leading to a net convective blueshift and characteristic line asymmetries \citep[e.g.][]{Dravinsetal1981,Dravins1982,Dravins1987,Asplundetal2000}. The dynamic nature of granulation produces temporal variations in line positions and profiles. On solar-like stars, these variations give rise to radial velocity (RV) signals with amplitudes ranging from several tens of cm\,s$^{-1}$ up to about 1\,m\,s$^{-1}$ \citep{Dumusqueetal2011,Dumusqueetal2015,Meunieretal2015,Milbourneetal2019,AlMoullaetal2023}, commonly referred to as granulation-induced RV jitter. RV amplitudes of this order are readily measurable by the new generation extreme-precision RV spectrographs \citep[see Figure~1 of][]{Crassetal2021,Johnetal2023}. Most importantly,  the RV signal induced by an Earth-like planet orbiting in the habitable zone of a Sun-like star is on the order of 10\,cm\,s$^{-1}$, i.e. smaller than the signal from granulation. Consequently,  granulation-driven RV jitter can mimic or even mask the planetary signals in the data from extreme-precision radial velocity searches for rocky planets. 

Efforts to understand and mitigate stellar granulation-induced RV jitter began in the early 2010s. They were based on observations, data-driven approaches, semi-empirical modeling and 3D numerical simulations of stellar granulation. Quantification of the granulation RV amplitudes using solar and stellar data led to the recommendation that at least 900\,s of integration of the signal and several measurements per night are needed to reduce granulation-driven RV jitter \citep{Dumusqueetal2011}. \cite{Bastienetal2014} proposed to use photometric flicker as a proxy for convective granulation as well as to use it to pre-select targets with intrinsically low granulation noise. HARPS-N \citep{Cosentinoetal2014} data enabled characterization of the RV jitter and its dependence on magnetic activity and spectral types \citep{Meunieretal2017a,Meunieretal2017b}. Large stellar samples later on refined empirical jitter-stellar parameter relations \citep[e.g.][]{Luhnetal2020}. Exploration of the spectral line shape changes in HARPS-N solar spectra demonstrated the potential of cross correlation function asymmetries as a diagnostic of convective blueshifts, enabling partial removal of granulation signals \citep{CollierCameronetal2019}.

Alongside these studies, data-driven mitigation methods based on principal component analysis \citep{CollierCameronetal2021} and Gaussian process regression \citep{Rajpauletal2015,Pergeretal2021,OSullivanetal2024} have been developed. These techniques are being refined with the help of continuous, high signal-to-noise datasets from modern Sun-as-a-star campaigns with HARPS-N and ESPRESSO \citep{CollierCameronetal2019,Milbourneetal2019} in order to minimize the granulation-induced RV jitter. In parallel, advances have been made in developing semi-empirical models to simulate solar RV \citep{Meunieretal2015,Palumboetal2022} and extend them to predict RV jitter for different stellar types \citep{Meunieretal2019,Dalaletal2023}, as well as in developing 3D numerical simulations of granulation for stars with different fundamental parameters \citep{Freytagetal2012,Allendreetal2013,Magicetal2013,Beecketal2013a,Beecketal2013, Bhatiaetal2022,Witzkeetal2023,Bhatiaetal2024}. 

One of the most promising directions for mitigating the granulation-induced RV jitter is to use differential effects in stellar spectra. The main idea here is that RV signals induced by planets are invariant from one spectral line to another across the entire spectrum, whereas granulation-induced signals are local to individual spectral lines. Therefore, measuring RVs on a line-by-line basis or in groups of lines with different sensitivities to granulation offers a means of disentangling planetary and stellar contributions. To this end, line-by-line RV extraction techniques which exploit the differential sensitivity of spectral lines to granulation are being developed \citep{Dumusque2018,Cretignieretal2020,AlMoullaetal2022}. The key ingredient needed for these techniques is the precise knowledge of differential sensitivity of spectral lines to granulation \citep{Dumusque2018}. Consequently, a comprehensive characterization of the sensitivity of all lines used in RV masks is urgently needed. 

This can now be achieved using ab initio 3D (magneto)hydrodynamic (MHD) models, which have reached a high degree of maturity and allow a quantitative characterization of convection. Using 3D MHD simulations with the MURaM code \citep{Voegleretal2004, Voegleretal2005}, \cite{Ceglaetal2013,Ceglaetal2018,Ceglaetal2019} developed parameterizations of granulation-driven signals for a single selected spectral line. \cite{Frameetal2025} used more recent MURaM simulations from \cite{Witzkeetal2024} and extended analysis to four spectral lines. \citet{DravinsandLudwig2023} investigated the temporal variability of the granulation RV jitter in spectra using hydrodynamic simulations with the CO5BOLD code \citep{Freytagetal2012}, focusing on the trends between weak/strong, neutral/ionized, low/high excitation energy, and atomic/molecular lines.

Here we develop a novel methodology to characterize signatures of granulation in spectral lines and understand the differential effects with the help of MURaM simulations of solar granulation by \cite{Witzkeetal2024}, which have been tested against quiet solar observations. We note that granulation is ergodic - which essentially means that the spatially averaged properties of granulation are equal to the temporally averaged properties, unless evaluated over scales smaller compared to the characteristic granulation length and lifetime. We exploit the ergodic nature of granulation and thus represent the sensitivity of spectral lines in the time domain by their variability across the stellar surface. In simple terms, spectral lines whose profiles and positions are similar across the stellar surface (e.g., in both granules and intergranular lanes) are not expected to exhibit strong temporal variability, whereas lines whose profiles and positions differ markedly between these regions are expected to show pronounced temporal variability. This ergodicity allows us to characterize granulation using a single simulation snapshot, thereby significantly reducing the computational costs. In this work, we focus on a relatively small sample of about hundred lines to study how the sensitivity of a line to granulation depends on its parameters and to explain the basic physics behind these established dependences. The behavior of all spectral lines used in RV masks, analyzed with a single snapshot of granulation and a methodology for mitigating the impact of granulation on RV measurements of solar twins, will be presented in our forthcoming study (Collier Cameron et al., in preparation). 

This paper is organized as follows. In \sref{sec:ergodicity} we present the methodology and use it to demonstrate the ergodic equivalence between the spatial and temporal variability of granulation-induced RV shifts in spectra. In \sref{sec:results-real} we apply the methodology to analyze sensitivities of Fe and Ti lines in our sample. The spectral line sample employed in the analysis is described in \sref{ssec:sample} while the metrics used to quantify the spatial variability are defined in \sref{ssec:metrics}. The response of Fe and Ti lines to granulation is presented and discussed in \sref{ssec:fetilines}. Using a set of artificial lines, hereafter referred to as \textit{sowmium}, we investigate how line parameters influence their formation and behavior. These results are presented in \sref{sec:results-sowmium}. Conclusions and an outlook are given in \sref{sec:conclu}. 

\begin{figure}
    \centering
    \includegraphics[scale=0.4]{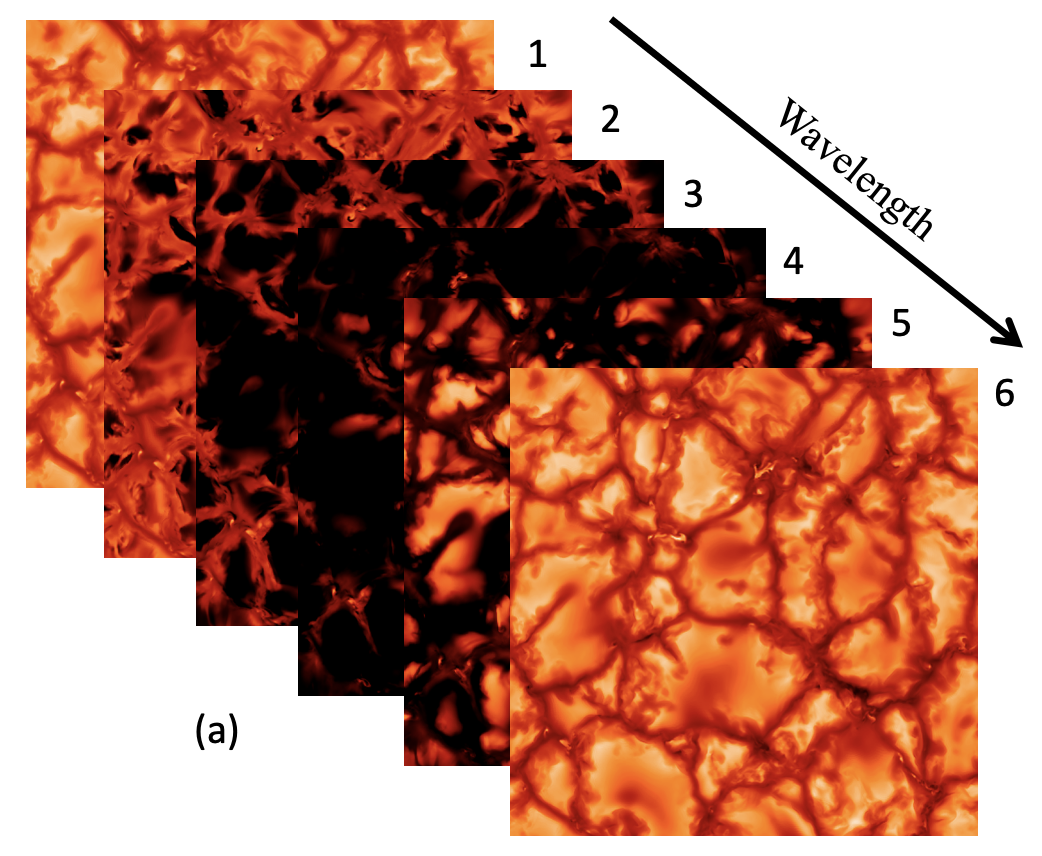}
    \includegraphics[scale=0.36]{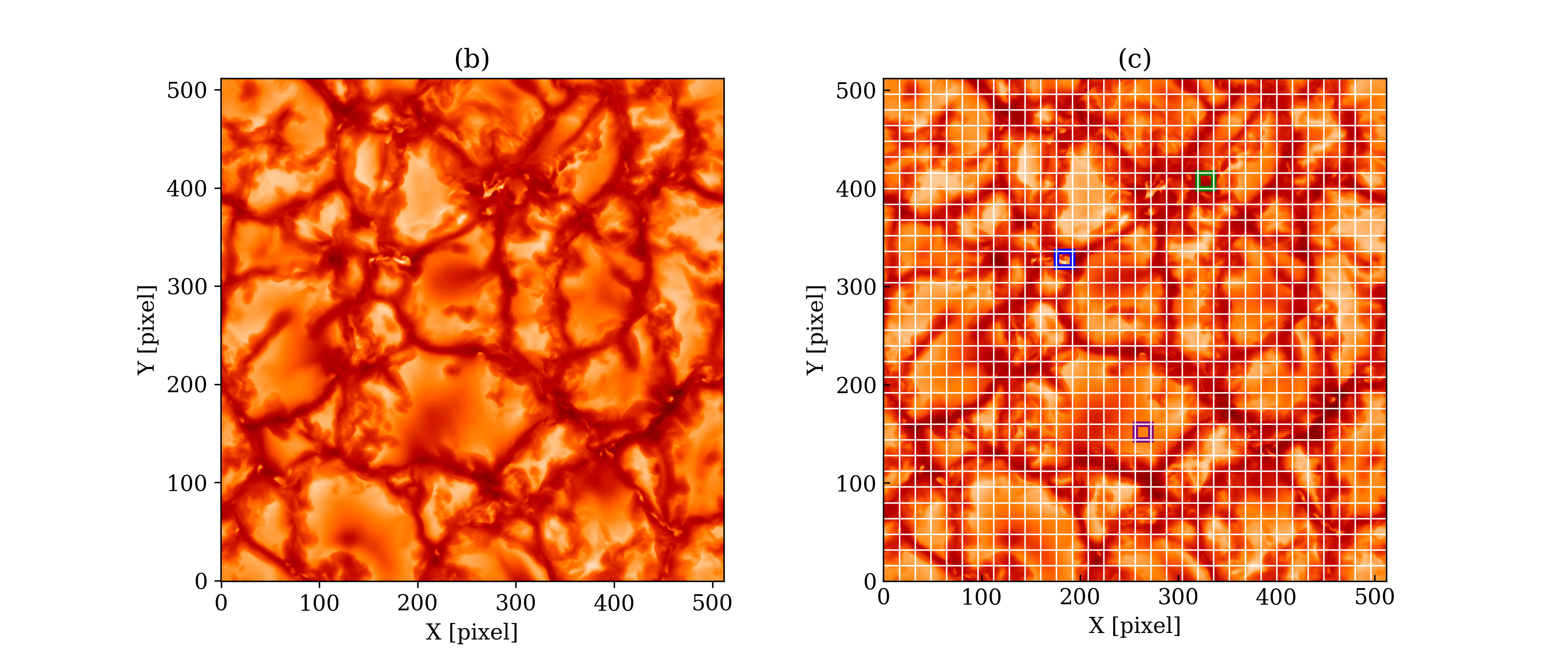}
    \includegraphics[scale=0.36]{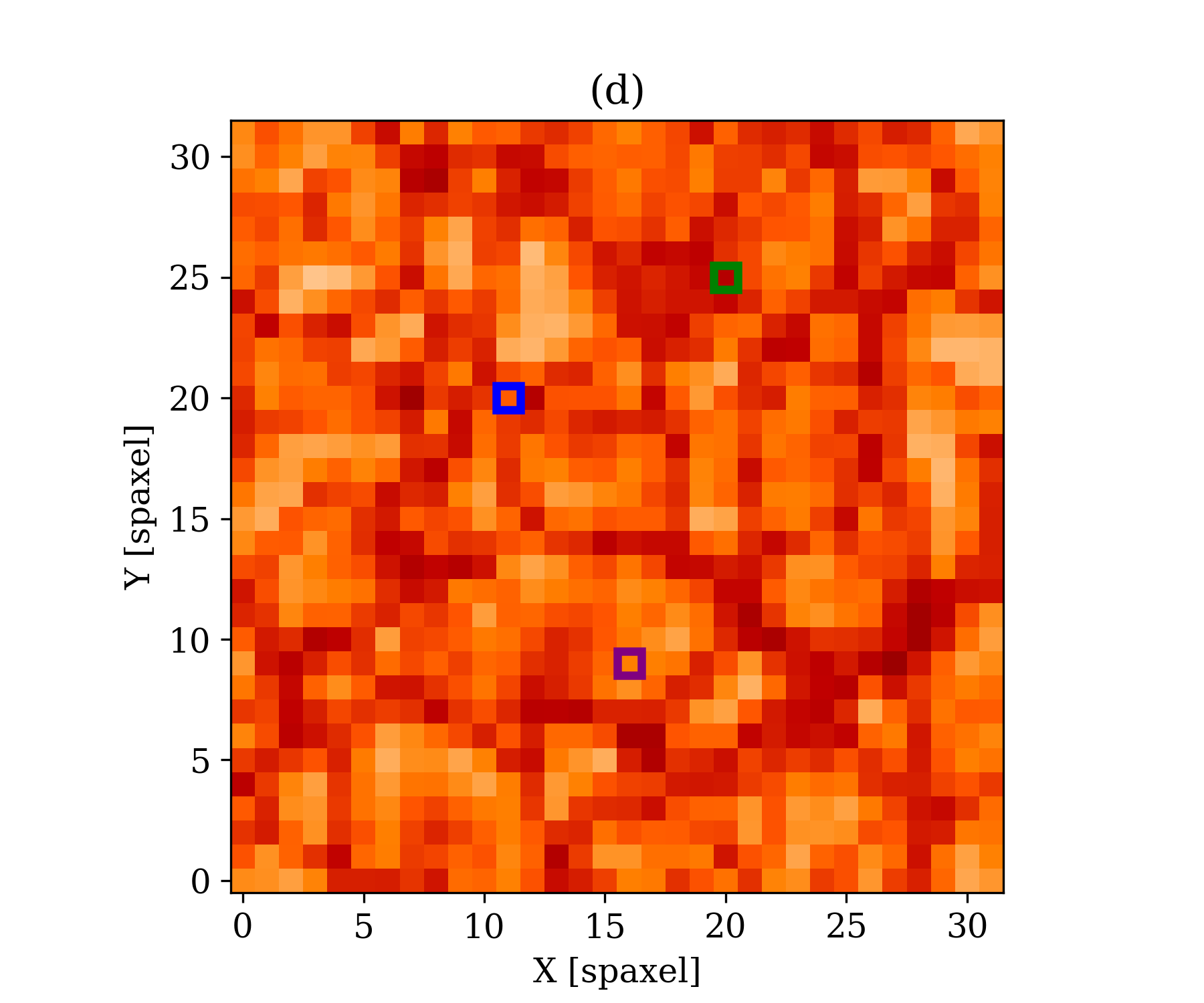}
    \includegraphics[scale=0.50]{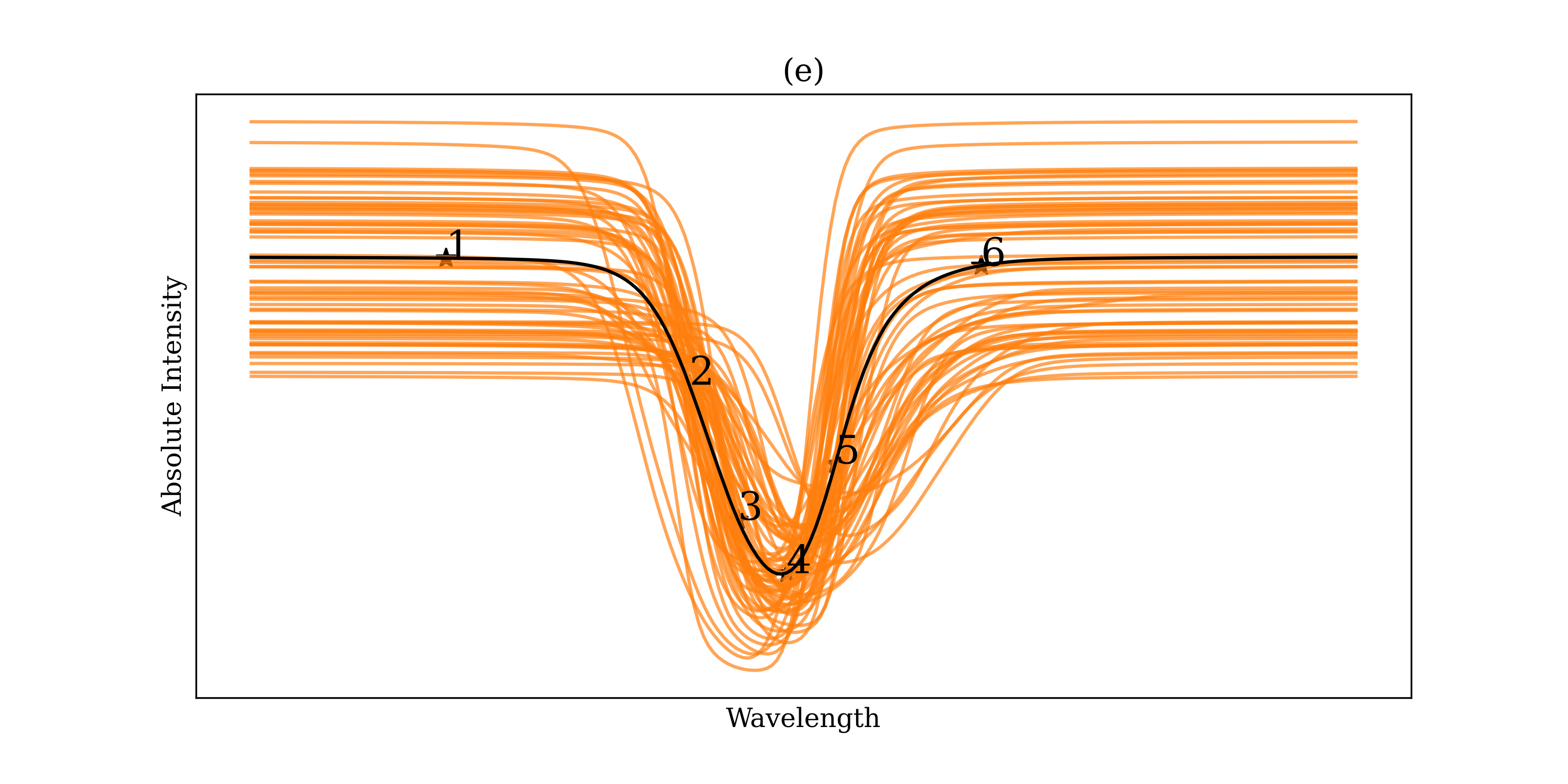}
    \caption{An illustration of the methodology used to analyze the spatial variability of spectral lines due to solar granulation. Panel a: intensity spatial maps at different wavelength points in the continuum and spectral line. These wavelengths are indicated by the numbers 1--6 in panel e. Such intensity maps form a 3D data cube, with two spatial dimensions and a wavelength dimension. Panel b: top view of the 3D data cube displaying intensity map at a continuum wavelength and at the original spatial resolution of the MURaM snapshot. Panel c: grid lines drawn on the map at the original spatial resolution to indicate the boundaries of the larger spatial elements, which we call as spaxels, into which the image will be decomposed. Panel d: intensity map at a low spatial resolution obtained after combining pixels to form spaxels. The three spaxels marked in blue, green and purple in panels (c) and (d) are discussed in \app{app:pixelsinspaxel}. Panel e: line profile from various spaxels in panel d (orange profiles) and the profile averaged over all spaxels (black profile). See \sref{ssec:setup} for details.}
    \label{fig:approach}
\end{figure}

\section{Methods}
\label{sec:ergodicity}
\subsection{The setup}
\label{ssec:setup}
To analyze the sensitivity of spectral lines to granulation, we use MURaM simulations of the solar surface granulation presented in \citet{Witzkeetal2024}, which have been shown to reproduce solar observables including spectral line shapes. These simulations are done following the so-called `box-in-a-star' approach, with a Cartesian box covering 9000\,km\,x\,9000\,km (512\,x\,512 pixel$^2$) in the horizontal direction with a resolution of $\sim17.57$\,km and 5000\,km in the vertical direction with a resolution of 10\,km. The horizontal extent of the simulation box is such that several tens of granules are clearly resolved at any point in time. A small-scale dynamo (SSD) operating in and below the near-surface layers and leading to the formation of small mixed polarity magnetic features is taken into account using the approach of \cite{Rempel2014,Rempel2018}. 

We note that the box-in-a-star approach does not allow direct calculations of absolute RV jitter and hence scaling from the box to the full stellar surface is necessary \citep[see e.g.][]{Shapiroetal2017}. Since we focus on the differential sensitivity of lines, we do not apply this scaling.

We calculate spectra using MURaM 3D atmospheres under the assumption of local thermodynamic equilibrium, with the MPS-ATLAS code \citep{Witzkeetal2021}, using the atomic parameter lists from the Vienna Atomic Line Database\footnote{https://vald.astro.uu.se/} \citep[see also][]{Piskunovetal1995,Ryabchikovaetal2015}. Radiative transfer calculations are carried out using the  ray-by-ray approach. Here we employ only the disk center ($\mu=1.0$, where $\mu={\rm cos}(\theta)$ and $\theta$ is the angle between the direction to the observer and the local solar surface normal) spectra for further investigations. The sensitivity of lines at other viewing angles is discussed in \app{app:viewingangle} where we argue that our disk center results are also representative for the disk-integrated case.

\fref{fig:approach}a shows specimen 2D intensity maps at different wavelengths across a spectral line obtained with the MPS-ATLAS code. This sequence of maps illustrates how the appearance of the granulation pattern changes across wavelengths sampling a spectral line and the adjacent continuum, as indicated in \fref{fig:approach}e. Together, these wavelength-dependent intensity maps form a 3D data cube (two spatial dimensions and one spectral dimension), where each spatial location is associated with a full emergent spectrum.

\fref{fig:approach}b displays the top view of the 3D intensity data cube at full numerical resolution (512\,x\,512 pixels). To reduce the spatial resolution and mimic observations with finite spatial sampling, we combine neighboring numerical pixels into larger spatial elements, which we hereafter refer to as spaxels (spatial pixels, a common terminology used in integral-field spectroscopy to define the smallest spatial unit for which an individual spectrum is available). The boundaries of these coarser spaxels are indicated by the white grid in \fref{fig:approach}c. For each spaxel, we compute the spatially averaged emergent spectrum by averaging the intensities of all pixels contained within that spaxel at every wavelength. The resulting lower-resolution intensity map at a representative wavelength is shown in \fref{fig:approach}d.

The main reason for introducing coarser spaxels rather than analyzing individual numerical pixels is that line profiles at the native grid scale often exhibit highly complex shapes, which makes their characterization using any metric less straightforward. Spatial averaging within spaxels produces smoother, more representative spectra as in \fref{fig:approach}e that are better suited for quantitative analysis.

\subsection{Main Idea of Our Approach: From Spatial to Temporal Variability}
Granulation on the solar surface produces variability both in time and across the stellar surface. This allows two complementary approaches: following a single spaxel in time to investigate the temporal evolution of its spectrum, or analyzing a single snapshot to examine how the emergent spectrum varies between spaxels. As a stochastic phenomenon, granulation is expected to be ergodic. Ergodicity implies that both approaches to quantifying variability should yield consistent results, provided that spatial variability is assessed over a sufficiently large number of spaxels and temporal variability is traced over sufficiently long time series for individual spaxels. The main requirement in both cases is that the spaxels capture all possible mixtures of surface phenomena caused by granulation (e.g. there are spaxels fully dominated by intergranular lanes and there are also spaxels dominated by granules). In other words, we expect that the distribution of spatial variations in granulation pattern at a single point in time to be similar to the distribution of variations in granulation over time.  This can be illustrated by considering spectral lines that have the same profile and positions in granules and intergranular lanes.  These lines are not expected to vary in time due to granulation. In contrast, lines which appear different in granules and intergranular lanes are expected to also show strong variability in time.

The main idea of our approach is to build on the ergodic nature of granulation and to estimate the sensitivity of spectral lines to granulation using spaxel-to-spaxel variability. Such spatial variability can be derived from only a few MURaM snapshots (indeed, we will show that even a single snapshot provides a reasonably accurate estimate; see Figure~\ref{fig:timevariation} in \app{app:singlesnap}). This allows us to avoid the computationally expensive spectral synthesis over time series spanning several stellar hours that would otherwise be required to characterize temporal variability. As a result our approach substantially reduces the computational costs. Furthermore, it eliminates the need to deal with the challenges arising in the temporal domain due to oscillations in the simulation box. These introduce a RV component that is not necessarily representative of true solar RV signal due to oscillations \citep[see][for a detailed discussion and the method to mitigate oscillation-driven RV variability in simulations]{Ceglaetal2013, Frameetal2025}. 

Throughout this study we work with spaxels, that are made by combining 16\,x\,16 pixels, leading to a 32\,x\,32 spaxels map as displayed in \fref{fig:approach}d. Each spaxel has an area of 280\,km\,x\,280\,km which is sufficient to resolve individual granules and intergranular lanes. A few examples of the average spectral line profiles from spaxels are shown in \fref{fig:approach}e, for a selected individual spectral line. A wide variety of line profile shapes and continuum intensities can be seen due to different mixtures of granular and intergranular pixels in different spaxels. See \app{app:pixelsinspaxel} for a few examples of line profiles from all pixels that make up a spaxel. Most importantly, in \app{app:spaxelsize} we demonstrate that the results of our analysis are independent of spaxel size.

\subsection{An Illustrative Test of Granulation Ergodicity}
\label{sec:ergodictest}
The main idea of this study is to establish a method to obtain fast and reliable estimates of spectral line sensitivity to granulation. Here we perform a simple illustrative test of ergodicity. For this test, we compute 32\,x\,32 spaxels for 45 MURaM snapshots at a cadence of 90\,s covering a total of 1.125 solar hours. We note that the spectrum for each spaxel covers 380--800\,nm at a spectral resolving power R=500000. The time variation can be represented as a stack of spectra of a single spaxel from a time series of 45 snapshots. We measure the RVs of those 45 spectra relative to the mean spectrum $\langle f \rangle$, derived by averaging the 1024 spaxels from any existing snapshot from the same model. To determine the relative Doppler shift of each spectrum in a time series compared to the mean spectrum, we compute the shift $\epsilon$ using the Taylor-series method of \citet{Bouchyetal2001}:
\begin{equation}
    \epsilon = -\frac{(f - \langle f \rangle) \cdot f'}{f' \cdot f'}.
    \label{eq:bouchyvel}
\end{equation}
Here $f'$ denotes the derivative of the spectral flux with respect to pixel velocity $dv=c\ d\lambda/\lambda$ and the dot products represent summation over all spectral pixels. The negative sign ensures that a positive Doppler shift corresponds to a positive rate of change of distance from the observer, consistent with a redshift. Both $(f - \langle f \rangle)$ and $f'$ are centered by subtracting their mean values before applying \eqn{eq:bouchyvel}.

\begin{figure}[ht]
    \centering
    \includegraphics[scale=0.5]{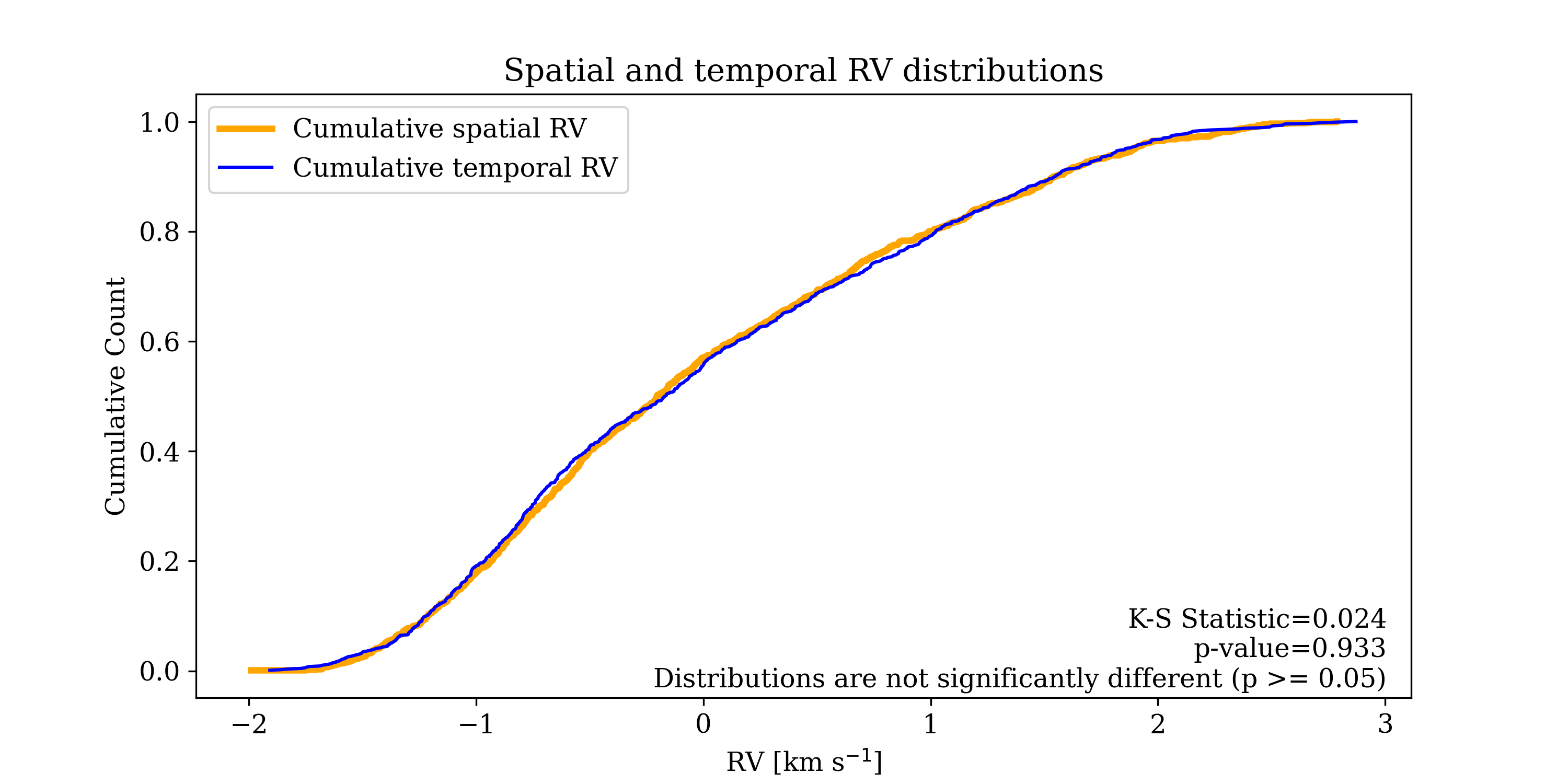}
    \caption{Result of the ergodicity test. Shown are the cumulative distributions of spatial RVs from a single snapshot and temporal RVs from 20 spaxels time series, computed using the methodology detailed in \sref{sec:ergodicity}. The two distributions are statistically indistinguishable, demonstrating the ergodic nature of the granulation driven RVs.}
    \label{fig:ergodicity}
\end{figure}

We extract 20 such spectral sequences from randomly-sampled locations in the snapshots and concatenate them to make a more densely sampled distribution. For the spatial variation, we compute the distribution of RVs for the spaxels in a single snapshot. The cumulative distributions of RVs from both calculations are essentially identical as shown in \fref{fig:ergodicity}. A Kolmogorov-Smirnov (K-S) test confirms that the two velocity distributions are effectively equivalent. This justifies our assumption that random sampling from a single snapshot captures the full range of behaviour that we expect to see in a long time series derived from any part of any snapshot. Therefore, in the rest of the paper, we focus on the spatial variations of granulation as seen in a single MURaM snapshot.

\section{Sensitivity analysis with solar lines}
\label{sec:results-real}
We will now apply the method developed in \sref{sec:ergodicity} to investigate the general trends in the sensitivities of spectral lines to variability. For illustration, we use a representative sample of about 100 spectral lines. In a forthcoming publication, we will extend these calculations to all spectral lines included in the HARPS-N mask and develop a methodology for mitigating the impact of granulation on RV measurements of solar twins.

\subsection{Spectral line sample}
\label{ssec:sample}
Our illustrative sample consists of spectral lines resulting from electronic transitions in the neutral and singly ionized states of Fe and Ti (Tables~\ref{tab:datafei}--\ref{tab:datatiii}). This choice is driven by the following considerations. Due to the complex electronic configuration of Fe atoms, numerous allowed transitions take place in them giving rise to millions of spectral lines in the visible range of the solar spectrum. As a result, the behavior of RV jitter in solar-like stars is to a large extent determined by the iron lines. While Ti lines are not as abundant as Fe lines, the first ionization potential of Ti (6.828 eV) is lower than that of Fe (7.902 eV). Thereby, Ti lines allow us to illustrate how ionization potential impacts line sensitivities to granulation. 

The line sample listed in Tables~\ref{tab:datafei}--\ref{tab:datatiii} cover a wide range of excitation energies of the lower level involved in the transition ($E_{\mathrm{low}}$) as well as of oscillator strengths (log($gf$)). The line sample was selected based on a visual inspection of the line strengths near the solar limb ($\mu=0.1$) intensity spectrum\footnote{https://www.irsol.usi.ch/it/data\_archive/ss2} \citep{Gandorfer2000,Gandorfer2002,Stenflo2005}. Our focus was to choose lines which are unblended (or at least not strongly blended given the high line density in the visible spectrum) and are clearly visible in the solar spectrum even at the limb in the wavelength range 430--530\,nm. We opted for this spectral range to illustrate the effects since it covers the window around 500\,nm where the solar irradiance spectrum per unit wavelength peaks. 

Convective velocities associated with granulation introduce net blueshifts in spectral lines, broaden the lines and make them asymmetric \citep{Dravinsetal1981}. Since our aim is to characterize the subtle changes in these granulation-driven effects, it is crucial to synthesize spectral lines at high spectral resolution. Therefore, we compute intensity profiles emerging at disk center for each of the lines in our sample at a spectral resolving power $R=2000000$. All in all, for each spectral line we perform calculations on 32 \,x\, 32 spaxels, disregarding possible blends, as they could obscure the general trends in line behavior that we aim to establish. 

\subsection{Metrics}
\label{ssec:metrics}
We begin by introducing formal definitions of center-of-gravity wavelength ($\lambda_\mathrm{cog}$) and equivalent width ($W_\lambda$) as follows
\begin{equation}
    \lambda_\mathrm{cog} = \frac{\sum_\lambda \bigg(1-\frac{I_\lambda}{I_c}\bigg) * \lambda}{\sum_\lambda \bigg(1-\frac{I_\lambda}{I_c}\bigg)} \,
    \label{eq:cog}
\end{equation}
and 
\begin{equation}
    W_\lambda = \sum_\lambda \bigg(1-\frac{I_\lambda}{I_c}\bigg) d\lambda\ .
    \label{eq:w}
\end{equation}
Here, $I_c$ is the intensity of the continuum and $I_\lambda$ is the intensity at a wavelength $\lambda$ in the spectral line. 

To analyze the sensitivity of spectral lines to granulation we focus on spaxel-to-spaxel variability of line strengths and central wavelengths. For any given spectral line we compute the two metrics in Equations~(\ref{eq:cog}) and~(\ref{eq:w}) for each of the 32\,x\,32 spaxels. Then we obtain their weighted means (which represent values computed for the emergent intensity from the entire cube) and weighted standard deviations as 
\begin{equation}
    <\lambda_\mathrm{cog}> = \frac{\sum_i \lambda^i_\mathrm{cog} * {w_i}}{\sum_i {w_i}} \ \ \ {\rm and}\ \ \ \sigma(\lambda_\mathrm{cog}) = \sqrt{\frac{\sum_i w_i (\lambda^i_\mathrm{cog}-<\lambda_\mathrm{cog}>)^2}{\frac{(N-1) \sum_i wi}{N}}}\ ,
    \label{eq:cogscatter}
\end{equation}
\begin{equation}
    <W_\lambda> =  \frac{\sum_i W^i_\lambda * {w_i}}{\sum_i {w_i}} \ \ \ {\rm and}\ \ \ \sigma (W_\lambda) = \sqrt{\frac{\sum_i w_i (W^i_\lambda-<W_\lambda>)^2}{\frac{(N-1) \sum_i wi}{N}}}
    \label{eq:ewscatter}
\end{equation}
The summation is performed over all spaxels indexed with $i$ which runs from 1 to $N$, where $N$ is the total number of spaxels within a single MURaM simulation snapshot ($=1024$). The weight $w_i$ of the $i$th spaxel is defined as $I^i_c/<I_c>$, the ratio of the continuum intensity in that spaxel to the continuum intensity computed from entire snapshot. Spaxels have a different mixture of granules, intergranules and at times bright points which form in the intergranular lanes (as seen in Figures~\ref{fig:approach}c and d). Some spaxels are dominated by granules while some others by intergranules. This results in both continuum intensity level and spectral line shape changes between spaxels as shown in \fref{fig:approach}e. The contribution from individual spaxel is therefore proportional to $I_c$ in that spaxel. For example, dark spaxels where $I_c$ is small contribute less to the resultant profile than bright spaxels with higher $I_c$. This is taken care by the weights $w_i$.

Lastly, we define line sensitivities to granulation as scatter in $\lambda_\mathrm{cog}$ and  $<W_\lambda>$ values normalized to their mean value, i.e.  $\sigma(\lambda_\mathrm{cog})/<\lambda_\mathrm{cog}>$ and $\sigma (W_\lambda)/<W_\lambda>$.

\begin{figure}
    \centering
    \includegraphics[scale=0.58]{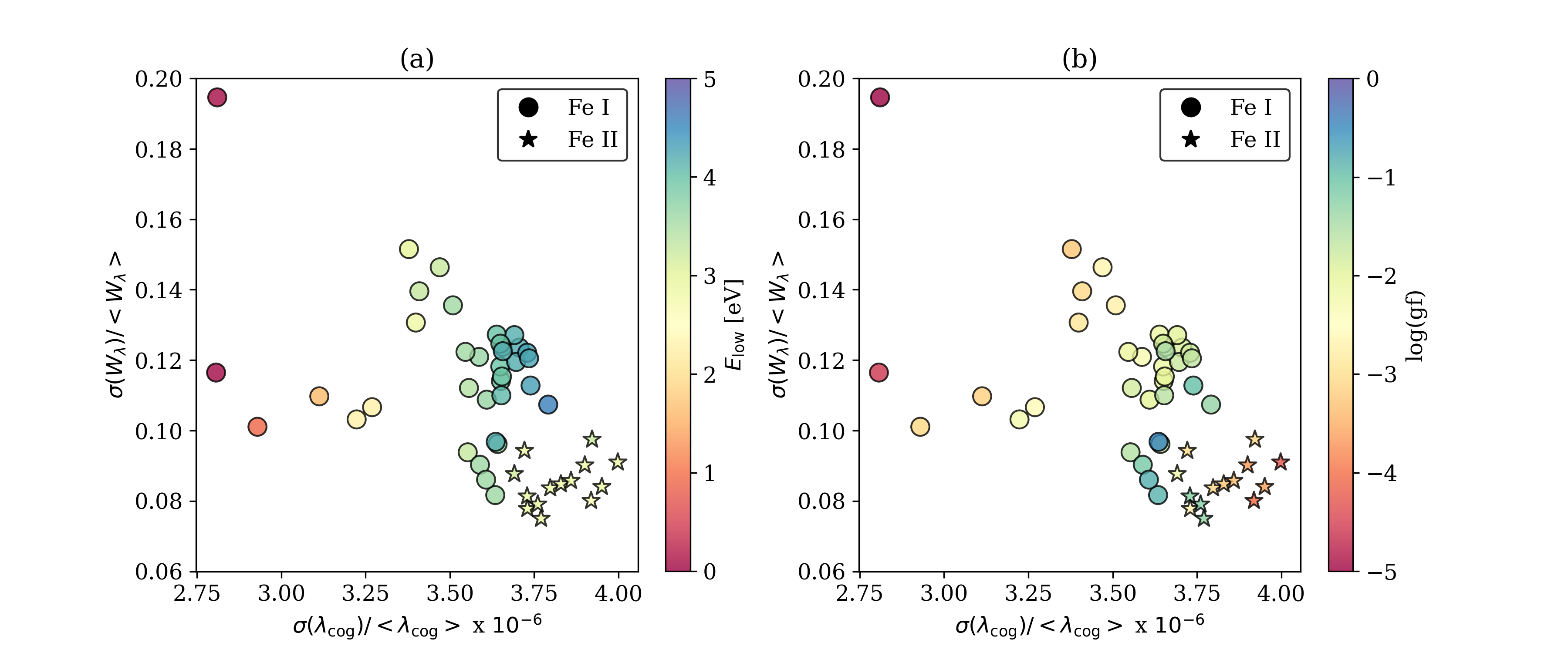}
    \caption{Differential sensitivity of spectral lines to solar granulation. Scatter in the line equivalent width ($\sigma W_\lambda$) normalized to the mean equivalent width ($<W_\lambda>$) is plotted as a function of the scatter in the center-of-gravity ($\sigma \lambda_\mathrm{cog}$) normalized to the mean center-of-gravity ($<\lambda_\mathrm{cog}>$). Fe\,I lines are indicated with filled circles and Fe\,II lines are shown as filled stars. The symbols are colored according to $E_\mathrm{low}$ values in panel (a) and according to log(gf) values in panel (b).}
    \label{fig:sensitivity}
\end{figure}

\begin{figure}
    \centering
    \includegraphics[scale=0.58]{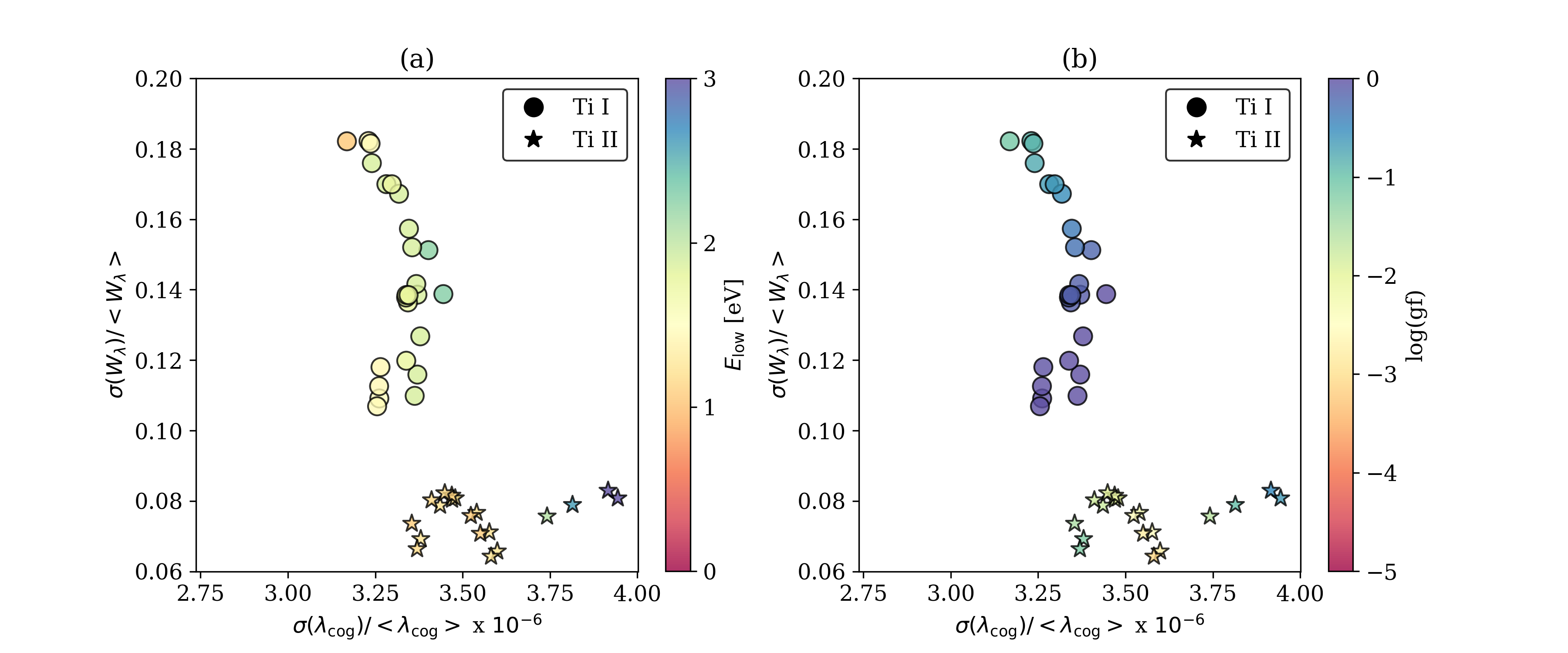}
    \caption{Same as \fref{fig:sensitivity} but now for Ti\,I and Ti\,II lines.}
    \label{fig:sensitivity_ti}
\end{figure}

\subsection{Behavior of Fe and Ti lines}
\label{ssec:fetilines}
Our main result is that line sensitivities to granulation vary strongly between individual lines. In this section we illustrate this for our sample of selected Fe and Ti lines (Figures~\ref{fig:sensitivity}~and~\ref{fig:sensitivity_ti}). We see from Figure~\ref{fig:sensitivity} that Fe\,I and Fe\,II lines form two distinct clouds of points, indicating that there is a clear difference in the way the two line species respond to granulation. While the strengths of Fe\,I lines show a stronger sensitivity (i.e., higher $\sigma (W_\lambda)/<W_\lambda>$) than Fe\,II lines, center-of-gravity of Fe\,II lines show a stronger sensitivity (i.e., higher $\sigma(\lambda_\mathrm{cog})/<\lambda_\mathrm{cog}>$) than those of Fe\,I lines \citep[similar to the findings of][]{DravinsandLudwig2023}. These trends are even more prominent in \fref{fig:sensitivity_ti} for the Ti\,I and Ti\,II lines.

We first explain why the strengths of Fe\,I and Ti\,I lines are generally more sensitive to granulation than those of Fe\,II and Ti\,II lines. For this we look closely into the solar photospheric plasma properties and spectral line formation characteristics. The temperatures and pressures in the solar photosphere are such that Fe and Ti atoms are mostly ionized resulting in the first ionization stage to be the dominant stage. The neutral Fe and Ti atoms become minority species. This implies that any spatial variations in the temperature because of the granulation pattern will have a far more significant impact on the populations of neutral Fe and Ti as compared to the populations of their singly ionized state. Such population changes in turn strongly impact strengths of neutral Fe and Ti explaining why the equivalent widths of their lines show greater variability across the granulation pattern.

Now we proceed with explaining why, in contrast to line strengths, the wavelength centers of gravity of Fe\,I and Ti\,I lines show lower sensitivity to granulation than those of Fe\,II and Ti\,II lines. An important factor that comes into play here is the variation in convective velocity amplitude across the granulation pattern in the line forming layers. In the solar photosphere, the convective velocities as well as the velocity difference between granules and intergranules decrease with height. The formation of Fe\,I and Ti\,I lines is restricted to cooler upper layers where the temperatures are not sufficient to ionize them. In these layers, convective velocities and the velocity contrast within the granulation pattern are smaller. Therefore the Fe\,I and Ti\,I lines show weaker sensitivity to the center-of-gravity wavelength displacements. On the contrary, the Fe\,II and Ti\,II lines form in hotter, deeper layers where both the convective velocities and their spatial variations across the granulation pattern are higher. This results in stronger sensitivity to center-of-gravity wavelength displacements for the Fe\,II and Ti\,II lines.

The separation in the sensitivity between neutral and inonized Ti lines seen in \fref{fig:sensitivity_ti} is more prominent than that for Fe seen in \fref{fig:sensitivity}. This is due to Ti having a lower ionization potential in relation to Fe. As a result Ti is more readily ionized than Fe in the solar atmosphere, making neutral Ti even more of a minority species than neutral Fe, thus amplifying both effects discussed above.

We note that the sensitivities shown in Figures~\ref{fig:sensitivity}~and~\ref{fig:sensitivity_ti} are calculated with a single MURaM snapshot. In \app{app:singlesnap}, we use a sample of 45 snapshots to demonstrate that calculations with a single snapshot provide an accurate estimate of the sensitivities.

Finally, we remark that no clear trends emerge when looking at the dependence of sensitivities on line parameters such as $E_\mathrm{low}$ and log(gf), although Fe\,I lines appear to show a strong dependence on $E_\mathrm{low}$ and lines from singly ionized elements with lower log(gf) appear to show higher sensitivity to center-of-gravity wavelength displacements. It is rather challenging to interpret the dependence of sensitivities on line parameters from Figures~\ref{fig:sensitivity} and~\ref{fig:sensitivity_ti}. This is because different effects due to line parameters such as $E_\mathrm{low}$, log(gf) and wavelength act simultaneously and make it difficult to isolate individual contributions. In order to investigate this closely and differentiate the multitude of effects, we work with a set of artificial lines for which the line parameters can be fixed manually. This sample and the results of its analysis are described in \sref{sec:results-sowmium}.

\begin{figure}
    \centering
    \includegraphics[scale=0.58]{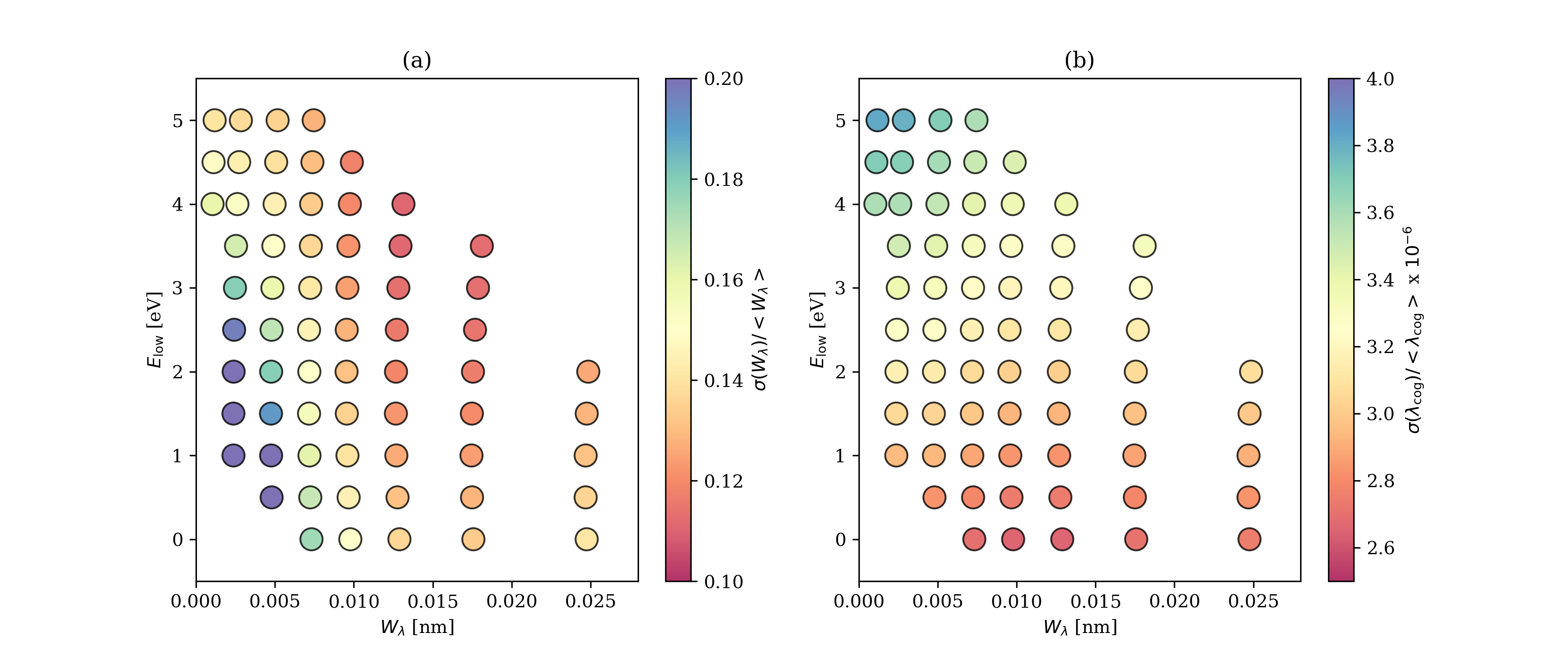}
    \caption{Granulation sensitivity of the sowmium neutral sample. The excitation energy of the lower level $E_\mathrm{low}$ is plotted as a function of the line equivalent width $W_\lambda$. Panel (a): points are color-coded according to the value of the variability in line equivalent width. Panel (b): the color-coding displays the variability in the line center-of-gravity wavelength displacements.}
    \label{fig:sowmiumi}
\end{figure}

\section{Dependence of line sensitivity on the line parameters}
\label{sec:results-sowmium}

\subsection{Sowmium sample}\label{subsec:sowmium}
To understand the dependence of line sensitivity on atomic parameters, we create an artificial spectral line sample by varying $E_\mathrm{low}$ and log(gf). We refer to this sample as `sowmium'. In essence, sowmium is the same as iron in all respects (atomic weight, ionization potential) but has a few extra fictional line transitions with properties that allow a more complete exploration of the parameter space. In this sample, there are 65 lines in the neutral state and 58 lines in the singly ionized state. Their wavelengths are fixed at 629.9532\,nm (the dependence of the granulation-driven line changes on wavelength of the lines in our sowmium sample is briefly discussed in \app{app:wavelength}). log(gf) is in the range 0 to -5 while $E_\mathrm{low}$ is in the range 0 to 5\,eV for the lines from neutral species and 2 to 6\,eV for the lines from singly ionized species, with a step size of 0.5 in both $E_\mathrm{low}$ and log(gf). $W_\lambda$ of these lines are in the range 0.001--0.025\,nm, which is the typical range into which most of the solar spectral lines fall. 

For further analysis, we focus on $E_\mathrm{low}$-$W_\lambda$ space instead of going to the $E_\mathrm{low}$-log(gf) space. This choice is driven by the fact that $W_\lambda$ increases monotonically with increasing log(gf) and lines with similar $W_\lambda$ form at similar heights in the solar atmosphere, sampling the same plasma properties. Due to this, we expect to see clearer trends in $E_\mathrm{low}$-$W_\lambda$ space.

Many trends emerge when looking at the dependence of sensitivities of neutral sowmium lines on $E_\mathrm{low}$ and $W_\lambda$ (see \fref{fig:sowmiumi}). First, we see that at a given $E_\mathrm{low}$, the sensitivity to the equivalent width increases strongly towards lines with smaller $W_\lambda$ i.e., for weaker lines (horizontal trend in \fref{fig:sowmiumi}a). Second, for lines with similar $W_\lambda$, the sensitivity to the equivalent width increases strongly towards lines with smaller $E_\mathrm{low}$ (vertical trend in \fref{fig:sowmiumi}a). Third, at a given $E_\mathrm{low}$, the sensitivity to line center-of-gravity displacements increases towards lines with smaller $W_\lambda$, however this increase is weak (horizontal trend in \fref{fig:sowmiumi}b) and fourth, for lines with similar $W_\lambda$, the sensitivity to the line center-of-gravity displacements becomes larger with increasing $E_\mathrm{low}$ (vertical trend in \fref{fig:sowmiumi}b).

\subsection{Dependences on equivalent width for neutral sowmium}\label{subsect:width}
\begin{figure}
    \centering
    \includegraphics[scale=0.27]{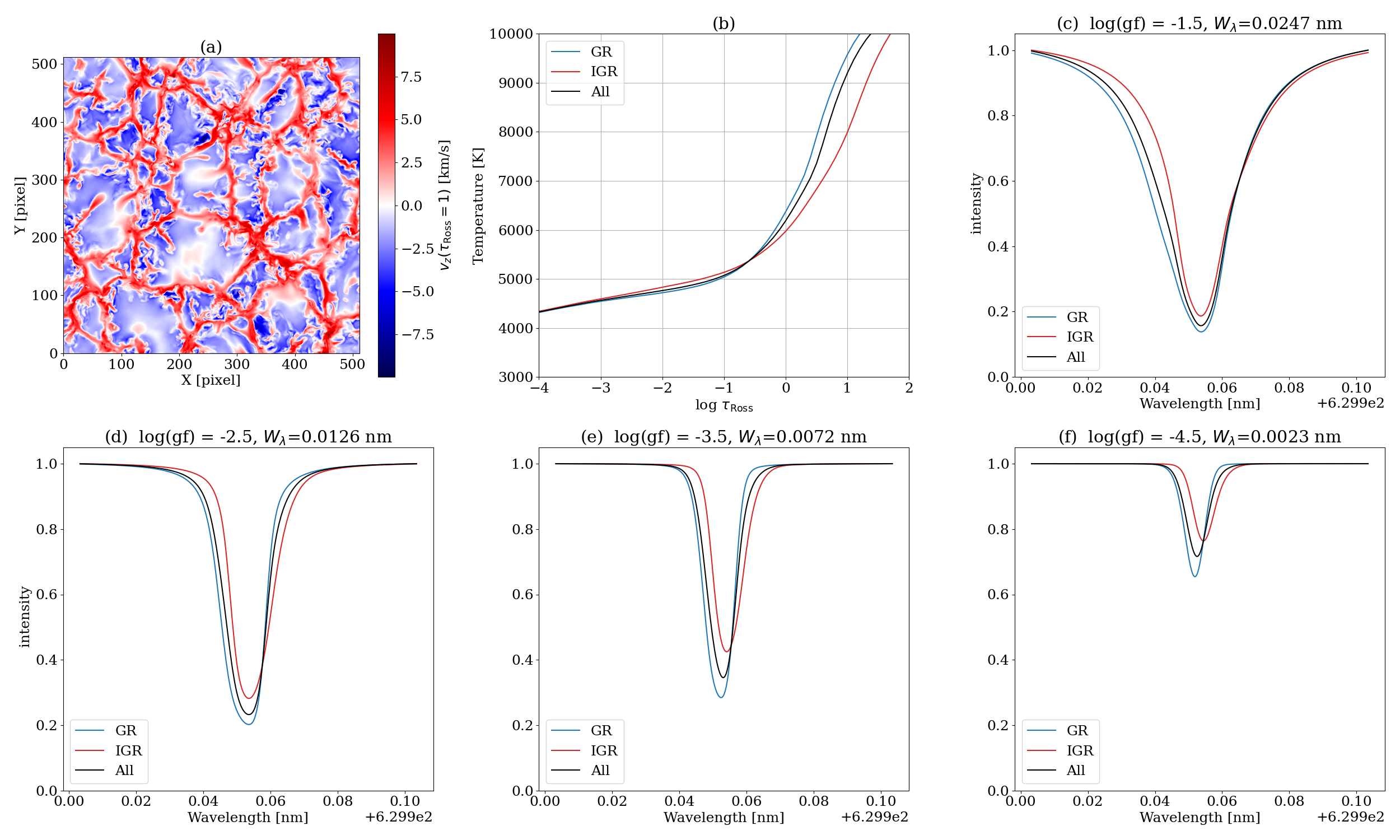}
    \caption{Panel (a): vertical velocities ($v_z$) from MURaM simulation at $\tau_\mathrm{Ross}=1$ layer. The sign has been chosen to follow the observational convention, such that negative velocities (blue) correspond to upflows (granules) and positive velocities (red) to downflows (intergranules). Panel (b): mean temperature structure of the granule (GR, blue), intergranule (IGR, red) and all pixels combined (All, black). Panel (c)-(f) normalized intensity profiles of neutral sowmium lines with $E_\mathrm{low}=1.5$\,eV and varying log(gf) and $W_\lambda$ values of which are indicated in the titles. See text for details.}
    \label{fig:fixedelowI}
\end{figure}

To explain the effects seen as a function of $W_\lambda$, we turn to the contrast between mean line profiles from granular and intergranular lanes. This is a simplified two component model of granulation which does not capture the true intricate pattern of granulation (which is properly accounted for by our spaxel approach) but it is sufficient to explain the trend with $W_\lambda$. For the purpose of identifying the granular and intergranular components, we use the vertical velocities ($v_z$) from a  MURaM simulation snapshot, at the layer where the Rosseland mean optical depth is unity. This velocity map is shown in \fref{fig:fixedelowI}a. 

Using this map, we create a mask by grouping all pixels with $v_z(\tau_\mathrm{Ross}=1)<0$ (blueshifts, towards the observer) into granules and pixels with $v_z(\tau_\mathrm{Ross}=1)>0$ (redshifts, away from the observer) to intergranules. Next, we apply this mask to the intensity 3D data cube (computed using the full velocity information from MURaM simulations) and obtain mean granular and intergranular line profiles.

We consider such two component profiles for four spectral lines which have the same value of $E_\mathrm{low}$ but different $W_\lambda$ (\fref{fig:fixedelowI}c-f). Stronger lines (with larger $W_\lambda$ values) form higher in the atmosphere where the temperature contrast between granules and intergranules is smaller (see left part of \fref{fig:fixedelowI}b). Further, these lines are more saturated. This leads to granular and intergranular profiles being nearly equally strong (see \fref{fig:fixedelowI}c) and hence the variability in line strength decreases. Weaker lines (with smaller $W_\lambda$ values and less saturation) form deeper in the atmosphere where there are large differences between temperature and temperature gradients in granules and intergranules (see left part of \fref{fig:fixedelowI}b). Steeper temperature gradients in the granules make the granular line profiles stronger, contributing more photons than the intergranules. This asymmetry in the strength of granular and intergranular components results in an increased sensitivity ($\sigma (W_\lambda)/<W_\lambda>$) of the line strength to changes in the granulation pattern.

At the same time, the sensitivities of line center-of-gravity displacements ($\sigma (\lambda_\mathrm{cog})/<\lambda_\mathrm{cog}>$) show only a weaker response to granulation variations. In order to explain this, we turn to the differences between vertical velocities in granules and intergranules and how they vary with height. When the line is strong, it forms in higher layers with smaller velocity contrast between granules and intergranules. Hence the line does not get shifted significantly. When the line is weak, it forms in deeper layers where the velocity contrasts are higher and hence the line should get significantly shifted. However, as discussed before, the granular line profile is much stronger than the intergranular line profile and determines the center-of-gravity wavelength. The strong asymmetry between the granular and intergranular profiles compensates for the stronger velocity shift. This effect is stronger than shown in \fref{fig:fixedelowI}, because the difference in continuum intensity between granules and intergranules gives added weight to the granular profiles. Therefore, the sensitivity of the line center-of-gravity displacements to granulation becomes less pronounced as the line gets weaker.

\subsection{Dependences on excitation energy for neutral sowmium}
\begin{figure}
    \centering
    \includegraphics[scale=0.27]{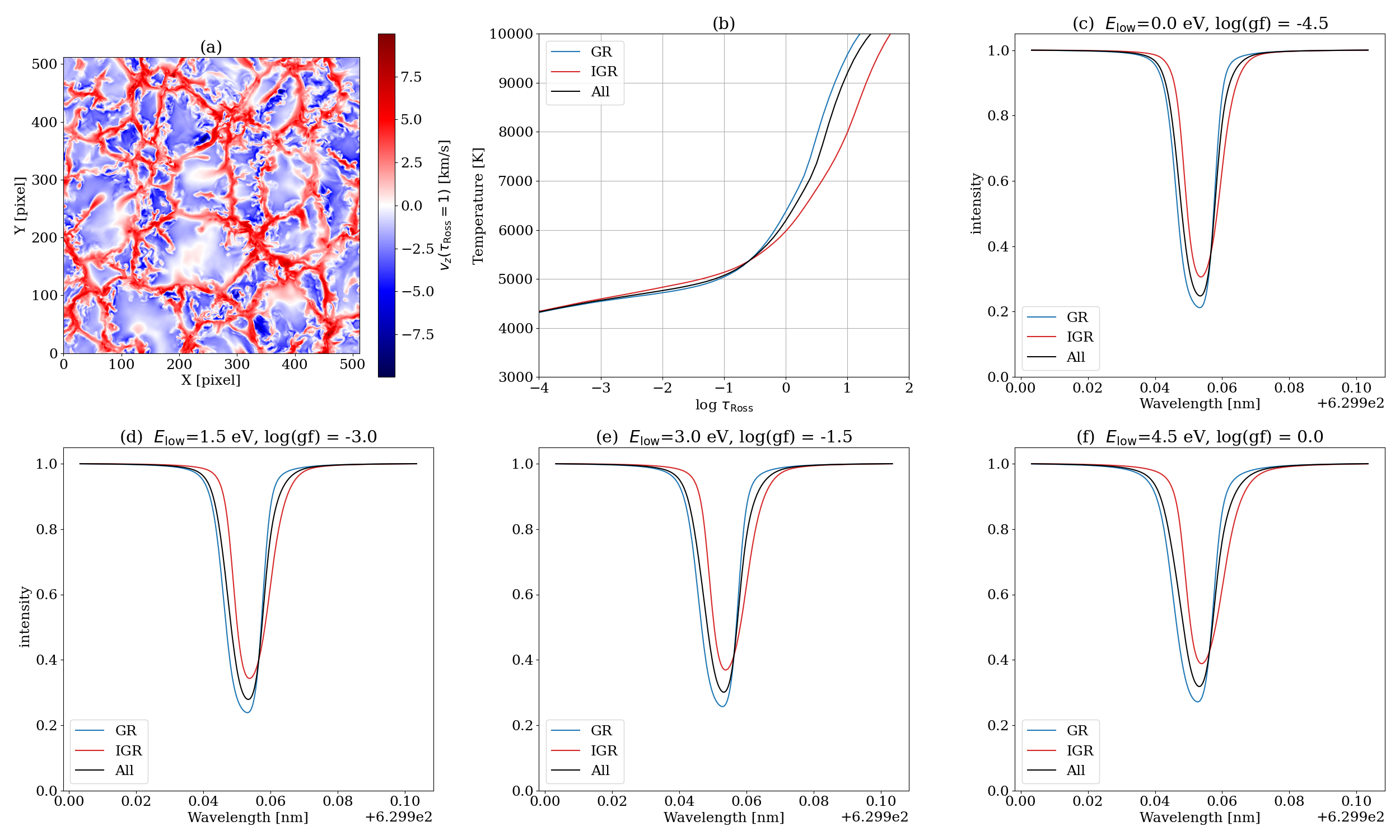}
    \caption{Panels (a)-(b) are same as in \fref{fig:fixedelowI}. Panels (c)-(f) now show neutral sowmium lines which have $W_\lambda\sim0.010$\,nm but with $E_\mathrm{low}$ and log(gf) values as shown in the panel titles. The blue, red, and black colors respectively represent granule, intergranule and all pixels combined.}
    \label{fig:fixedewI}
\end{figure}

In this subsection we focus on lines with the same equivalent width but different excitation energies. The mean profiles from granular, intergranular pixels and all pixels combined are shown in \fref{fig:fixedewI}c-f for four lines with similar $W_\lambda$. With decreasing $E_\mathrm{low}$, \fref{fig:fixedewI} presents no visibly distinct changes in the granular and intergranular profiles, making it difficult to interpret the decrease in the sensitivity to line strength at higher values of $E_\mathrm{low}$. 

The reason for this behavior is that lines with the same equivalent width have similar (though not identical) formation heights.
As a result the different sensitivities of these lines to granulation cannot be explained by the dependence of the contrast between granules and intergranules on formation height as we did in Sect.~\ref{subsect:width}. Thereby, a simplified explanation that relies on averages of granular and intergranular spectra is not anymore applicable. Instead we need a more detailed consideration that accounts for the full and complex pattern of granulation. 

\begin{figure}
    \centering
    \includegraphics[scale=0.5]{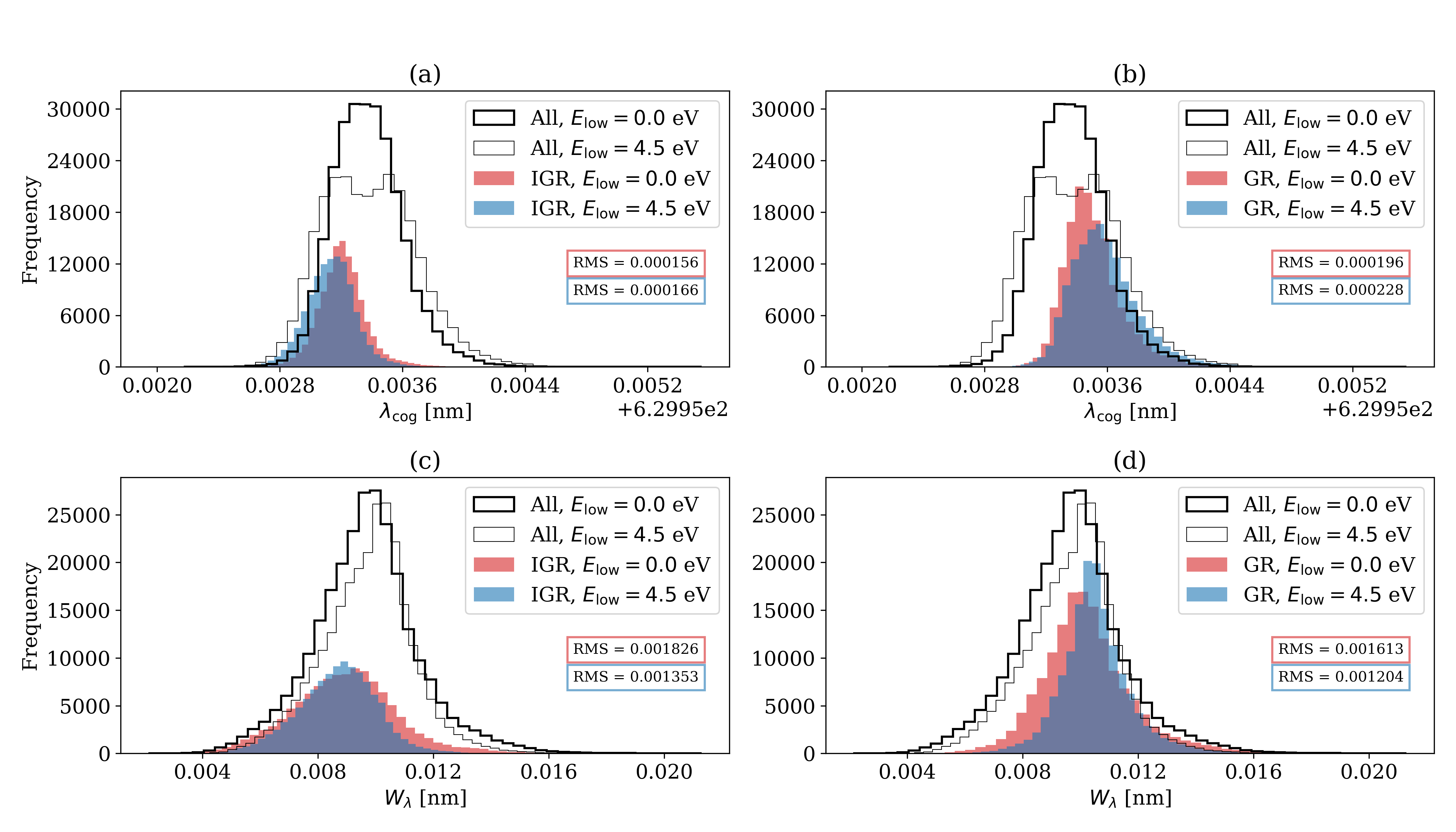}
    \caption{Distributions of line center-of-gravity wavelength displacements (top row) and line equivalent widths (bottom row) calculated for two neutral sowmium lines with $E_\mathrm{low}$ values as indicated in the panels. Thin and thick black histograms show the distributions for all 512\,x\,512 pixels in the MURaM simulation snapshot. The colored histograms show the distributions for only pixels within intergranular lanes (IGR, left column) and only pixels within granules (GR, right column). The RMS scatter of the corresponding distributions are also indicated. See text for further details.}
    \label{fig:hist-si-ver}
\end{figure}

We now look at the distribution of $\lambda_\mathrm{cog}$ and $W_\lambda$ within all granular and intergranular pixels for the two neutral sowmium lines shown in \fref{fig:fixedewI}c and \fref{fig:fixedewI}f. These distributions are shown in \fref{fig:hist-si-ver}. The root-mean-squared (RMS) values indicated in \fref{fig:hist-si-ver} clearly show that sensitivity to line center-of-gravity increases for lines with higher $E_\mathrm{low}$ while at the same time the sensitivity to strength decreases.

\subsection{Singly ionized sowmium lines}
\begin{figure}
    \centering
    \includegraphics[scale=0.58]{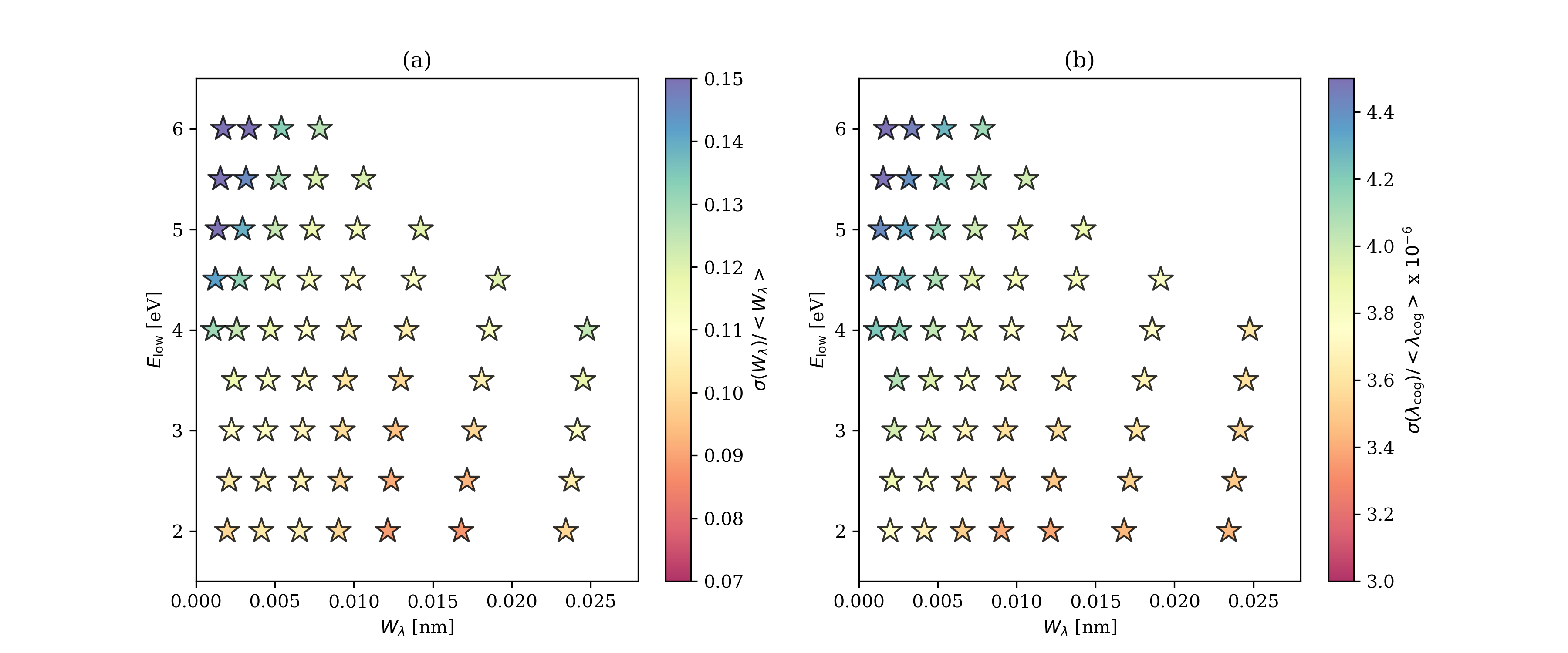}
    \caption{Same as \fref{fig:sowmiumi} but now for the sowmium singly ionized sample.}
    \label{fig:sowmiumii}
\end{figure}

Singly ionized sowmium lines, unlike neutral sowmium lines, show similar trends in line equivalent width and line center-of-gravity variability (see \fref{fig:sowmiumii}). The reason for this is a weaker (relative to the neutral case) effect of temperature difference between granules and intergranules on line strength for the ionized case (since sowmium is mostly ionized, see discussion for the iron and titanium cases in \sref{ssec:fetilines}). This weakens the horizontal trend in \fref{fig:sowmiumii}a but amplifies the horizontal trend in \fref{fig:sowmiumii}b, both with respect to neutral lines.

The sensitivity of line strength to temperature changes increases towards higher $E_\mathrm{low}$. This is because most ionized sowmium remains in the lower excitation stage, independent of temperature. By contrast, ionized sowmium with high excitation energy is preferentially found at higher temperatures, and thus shows a strong temperature sensitivity opposite to that of neutral sowmium. Thereby, both sensitivities shown in \fref{fig:sowmiumii}a and b increases with $E_\mathrm{low}$.

\section{Summary and outlook}
\label{sec:conclu}
Granulation-induced RV jitter is an obstacle for detecting Earth-like planets using extreme precision RV measurements. Reliable models of stellar granulation and effective mitigation techniques are needed to overcome this obstacle. To this end, we developed an approach to study and characterize spectral line variability arising from granulation and applied it to the Sun. Taking advantage of the ergodic nature of granulation, we opted to work with a single MURaM snapshot and analyze the spatial variability in the spectral lines. This allows us not only to make the calculations of the granulation-induced spectral line variability for the Sun and its extensions to other stars faster but also to circumvent effects due to acoustic oscillations in the temporal domain.

We limited our investigation to analyzing the response of the neutral and singly ionized iron and titanium lines to solar granulation. We found that the lines from neutral and ionized elements display very different sensitivities. The equivalent widths of lines of neutral species, and the center-of-gravity wavelength displacements in lines of ionized species, respond most strongly to contrasts in temperature and fluid velocity across the granulation pattern. Conversely, the equivalent widths of lines from ionized species and the center-of-gravity wavelength displacements of lines from neutral species show less sensitivity to granulation. These differences become more pronounced in titanium than in iron owing to the lower ionization potential of titanium. Further, by using a sample of artificial lines, we investigated the dependence of the line variability arising from granulation on the line parameters themselves. This analysis clarified how (and why) the sensitivity of line equivalent width and wavelength shift depends on line parameters (equivalent width and excitation energy of the lower level).

The approach described in this study and the investigations that were carried out form a first step towards identifying different classes of line variability resulting from granulation to enable careful line selections to effectively mitigate granulation-driven RV jitter. In the upcoming studies, we apply the novel method described here, based on the spatial variability of granulation, to analyze sun-as-a-star full visible solar spectrum from MURaM (Collier Cameron et al., in preparation) and to develop data-driven methods for identifying families of spectral lines showing the different types of response investigated here through stellar-scale sampling of the granulation pattern (Hartogh et al, in preparation).

\nolinenumbers
\begin{acknowledgments}
We acknowledge support from the European Research Council (ERC) under the European Union’s Horizon 2020 research and innovation program (grant No. 101118581 and No. 101097844). VV acknowledges support from the Max Planck Society under the grant ``PLATO Science'' and from the German Aerospace Center under ``PLATO Data Center'' grant   50OO1501. This study has made use of SAO/NASA Astrophysics Data System's bibliographic services.
We gratefully acknowledge the computational resources provided by the Raven supercomputer systems of the Max Planck Computing and Data Facility (MPCDF) in Garching, Germany.
\end{acknowledgments}

\software{
matplotlib \citep{matplotlib}, 
numpy \citep{numpy}
}

\appendix 
\restartappendixnumbering

\section{Profiles from pixels in spaxels}
\label{app:pixelsinspaxel}
\begin{figure}
    \centering
    \includegraphics[scale=0.45]{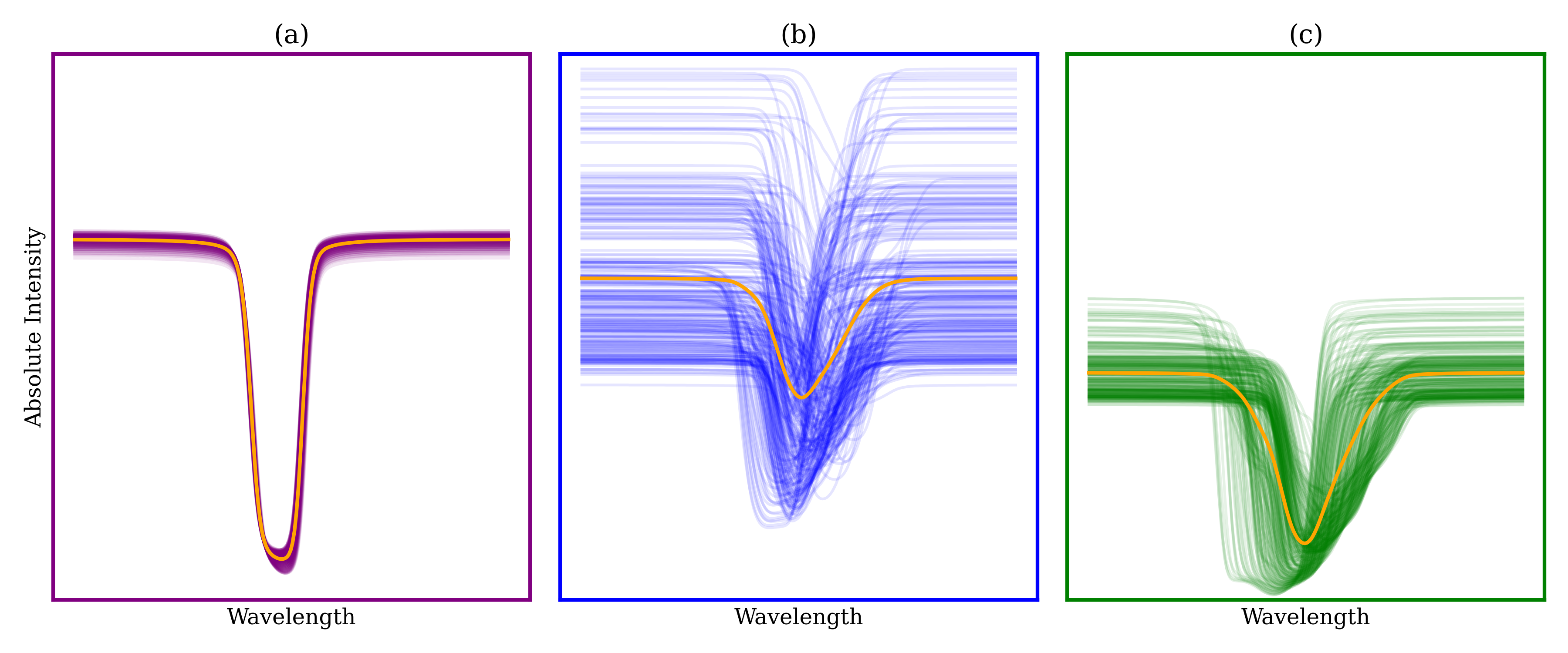}
    \caption{Examples of line profiles from 16\,x\,16 individual pixels constituting the mean spaxel profile (orange), shown for the three spaxels marked in \fref{fig:approach}. The profiles are plotted for a spaxel fully covered by granule (panel a), a spaxel with a mix of bright point and intergranular lane (panel b), and a spaxel covering the edge of a granule and intergranular lane (panel c). The profiles have been drawn on the same absolute intensity and wavelength scales for easy comparison.}
    \label{fig:pixelsinspaxel}
\end{figure}

The complexity of line profiles resulting from granulation is illustrated in \fref{fig:pixelsinspaxel}. Variations in line shapes, strengths and continuum intensities from one spatial pixel to the next are clearly visible. Line profiles shown in panel (a) originate from a spaxel which is fully covered by bright granular pixels (see \fref{fig:approach}) which are at a higher temperature compared to the intergranules and hence the continuum intensities are higher. The steep temperature gradients within granule make the line profiles deep (strong). Variations in the line profiles are small due to plasma properties being very similar across all pixels, resulting in the mean spaxel profile which is very similar to the constituent pixel profiles. 

The line profiles shown in \fref{fig:pixelsinspaxel}b originate from a spaxel having a mix of a magnetic bright point (a consequence of SSD) and dark intergranular lanes. As a consequence of non-thermal heating, the magnetic bright point is brighter than typical granules and hence the line profiles from magnetic bright point pixels lie above the level of granular profiles in \fref{fig:pixelsinspaxel}a. Line profiles from pixels which fall in dark intergranular lanes show lower continuum and line intensities as expected. The temperature gradients within intergranular lanes, where also the magnetic bright point forms, are not steep and hence the line profiles are shallow relative to those from granules. The mean spaxel profile is clearly different from the individual profiles comprising it but the line still displays a residual distortion due to the mix of  temperatures and velocities produced by granulation.

\fref{fig:pixelsinspaxel}c depicts line profiles from a spaxel that covers the edge of a granule (which in this case is darker than the granule center plotted in \fref{fig:pixelsinspaxel}a) and a dark intergranular lane. Here, the plasma rising at the edge of the granule interacts with the plasma falling back along intergranular lanes and can lead to shocks \citep{Cattaneoetal1990,Solankietal1996}. These interactions give rise to line profiles with shapes which are more complex than in \fref{fig:pixelsinspaxel}a. Smaller temperatures and temperature gradients in these spatial pixels give rise to shallow profiles and reduced continuum intensities as discussed before. Though the mean spaxel profile appears different, line distortions are clearly visible. 

This demonstrates that a group of spaxels which resolve a few dozen granules, capture intricate line shape changes due to granulation reasonably well, while averaging out the more extreme line profiles. 

\section{Fe and Ti line sample}
Tables~\ref{tab:datafei}--\ref{tab:datatiii} show the wavelength, $E_\mathrm{low}$ and log(gf) of the spectral lines from neutral and ionized Fe and Ti.
\label{app:sample}
\begin{table*}[htpb]
\centering
\caption{List of Fe I lines synthesized with \mpsa. The atomic parameters are taken from VALD3 database.}
\begin{tabular}{ccc}
\midrule[1.5pt]
\midrule
Vacuum wavelength & $E_\mathrm{low}$ & \multirow{2}{*}{log(gf)} \\
(nm) & (eV) & \\
\midrule[1.5pt]
439.0478 & 0.052 & -4.583\\ 
442.5082 & 3.651 & -1.610\\
444.6718 & 0.087 & -5.441\\
448.5477 & 3.601 & -0.864\\
452.4666 & 3.651 & -1.990\\
455.2922 & 3.939 & -2.060\\
458.4223 & 2.840 & -2.879\\
458.9006 & 3.975 & -2.150\\
459.4812 & 3.939 & -2.060\\
459.7702 & 3.651 & -2.320\\
459.9404 & 3.276 & -1.570\\
460.3289 & 1.605 & -3.154\\
462.0580 & 3.601 & -1.120\\
462.8842 & 3.229 & -3.059\\
463.1416 & 2.277 & -2.587\\
468.0154 & 3.601 & -0.833\\
472.7457 & 2.995 & -3.250\\
473.0340 & 4.068 & -1.614\\
475.1275 & 4.554 & -1.340\\
481.1282 & 3.568 & -2.720\\
490.6501 & 3.921 & -2.050\\
490.9102 & 3.423 & -1.840\\
491.1696 & 4.182 & -0.461\\
492.6143 & 2.277 & -2.241\\
492.8793 & 3.568 & -2.073\\
499.3259 & 4.211 & -1.910\\
499.5522 & 0.915 & -3.080\\
500.0506 & 4.182 & -1.740\\
501.4092 & 4.279 & -1.790\\
501.7877 & 4.250 & -1.690\\
502.1129 & 3.975 & -2.126\\
510.5612 & 4.172 & -1.970\\
511.1076 & 4.299 & -0.980\\
513.1059 & 3.939 & -1.850\\
523.7660 & 4.182 & -1.497\\
526.4345 & 3.246 & -2.660\\
\midrule[1.5pt]
\end{tabular}
\label{tab:datafei}
\end{table*}

\begin{table*}[htpb]
\centering
\caption{List of Fe II lines synthesized with \mpsa. The atomic parameters are taken from VALD3 database.}
\begin{tabular}{ccc}
\midrule[1.5pt]
\midrule
Vacuum wavelength & $E_\mathrm{low}$ & \multirow{2}{*}{log(gf)} \\ (nm) & (eV) & \\
\midrule[1.5pt]
449.2660 & 2.853 & -2.700 \\
458.4119 & 2.840 & -3.090 \\
462.1807 & 2.827 & -3.240 \\
465.8284 & 2.886 & -3.610 \\
466.8050 & 2.827 & -3.368 \\
467.1477 & 2.578 & -4.059 \\
492.5296 & 2.886 & -1.320 \\
499.4748 & 2.801 & -3.640\\
501.9834 & 2.886 & -1.220 \\
513.8218 & 2.840 & -4.290 \\
517.0467 & 2.886 & -1.250 \\
523.6080 & 3.216 & -2.230 \\
526.6266 & 3.224 & -3.120 \\
528.5562 & 2.886 & -2.990 \\
\midrule[1.5pt]
\end{tabular}
\label{tab:datafeii}
\end{table*}

\begin{table*}[htpb]
\centering
\caption{List of Ti I lines synthesized with \mpsa. The atomic parameters are taken from VALD3 database.}
\begin{tabular}{ccc}
\midrule[1.5pt]
\midrule
Vacuum wavelength & $E_\mathrm{low}$ & \multirow{2}{*}{log(gf)} \\ (nm) & (eV) & \\
\midrule[1.5pt]
439.5150 & 2.267 & +0.060\\
440.5518 & 2.246 & -0.954\\
440.5623 & 1.053 & -1.400\\
440.6126 & 1.877 & -0.670\\
440.6918 & 1.053 & -1.920\\
441.8506 & 1.885 & -0.100\\
442.2989 & 2.235 & -0.200\\
442.4059 & 1.065 & -1.100\\
442.7290 & 1.877 & -0.310\\
442.8336 & 1.501 & +0.230\\
443.5258 & 1.872 & -0.146\\
444.1587 & 1.872 & -0.390\\
444.2511 & 1.872 & -0.800\\
445.0390 & 1.885 & +0.470\\
445.2144 & 1.877 & +0.320\\
445.4558 & 1.427 & -0.030\\
445.4949 & 1.872 & +0.100\\
445.6562 & 1.440 & +0.100\\
445.8676 & 1.457 & +0.260\\
446.4779 & 1.885 & -0.560\\
446.7056 & 1.735 & -0.130\\
447.2492 & 1.731 & -0.150\\
447.6098 & 1.440 & -0.910\\
448.0953 & 1.731 & -0.630\\
448.2512 & 1.747 & +0.170\\
448.3938 & 1.457 & -0.890\\
449.0346 & 1.735 & -0.150\\
449.7401 & 1.747 & -0.140\\
449.7491 & 0.021 & -2.620\\
\midrule[1.5pt]
\end{tabular}
\label{tab:datatii}
\end{table*}

\begin{table*}[htpb]
\centering
\caption{List of Ti II lines synthesized with \mpsa. The atomic parameters are taken from VALD3 database.}
\begin{tabular}{ccc}
\midrule[1.5pt]
\midrule
Vacuum wavelength & $E_\mathrm{low}$ & \multirow{2}{*}{log(gf)} \\
(nm) & (eV) & \\
\midrule[1.5pt]
438.8071 & 2.596 & -0.960\\
439.2259 & 1.228 & -2.300\\
439.5291 & 1.220 & -1.770\\
439.6258 & 1.082 & -0.540\\
439.7067 & 1.243 & -1.930\\
440.1000 & 1.237 & -1.200\\
441.0472 & 1.243 & -2.780\\
441.0755 & 1.228 & -2.530\\
441.2308 & 3.093 & -0.650\\
441.8949 & 1.162 & -1.190\\
441.9567 & 1.237 & -1.990\\
442.3175 & 2.058 & -1.640\\
443.3335 & 1.237 & -3.080\\
444.5044 & 1.080 & -0.710\\
444.5799 & 1.115 & -2.200\\
445.1726 & 1.082 & -1.520\\
446.5699 & 1.160 & -1.810\\
446.9747 & 1.131 & -0.630\\
447.0399 & 1.082 & -2.550\\
447.2108 & 1.162 & -2.020\\
448.9583 & 3.122 & -0.500\\
449.4776 & 1.080 & -2.780\\
\midrule[1.5pt]
\end{tabular}
\label{tab:datatiii}
\end{table*}

\section{Effect of spaxel size}
\label{app:spaxelsize}
\setcounter{figure}{0}
In this section, using Fe\,I and Fe\,II lines, we show that the spectral line sensitivities to granulation do not depend on the size of spaxels. The line sensitivity metrics similar to \fref{fig:sensitivity} are plotted for three different sizes of spaxels in \fref{fig:spaxelsize}. With increasing size of the spaxel, the sensitivities reduce in amplitude. However, the distinctly different behavior of Fe\,I and Fe\,II lines remains unaffected. The two species populate two different parts of this line variability diagram, irrespective of the spaxel size.

This behavior is also evident in the scatter plots shown in \fref{fig:spaxelsize_scatter}. In panels (a)-(c), we see that the sensitivity to the center-of-gravity wavelength displacements for different spaxel sizes is linearly correlated. The red points lie above the blue ones, showcasing higher sensitivity in center-of-gravity wavelength displacement of Fe\,II lines. The sensitivity to the equivalent width is also linearly correlated at different spaxel sizes as seen in panels (d)-(f). Here, the blue points lie above the red ones, since Fe\,I lines higher sensitivity to equivalent width. See \sref{sec:results-real} for a more detailed discussion of this behavior.

\begin{figure}
    \centering
    \includegraphics[scale=0.38]{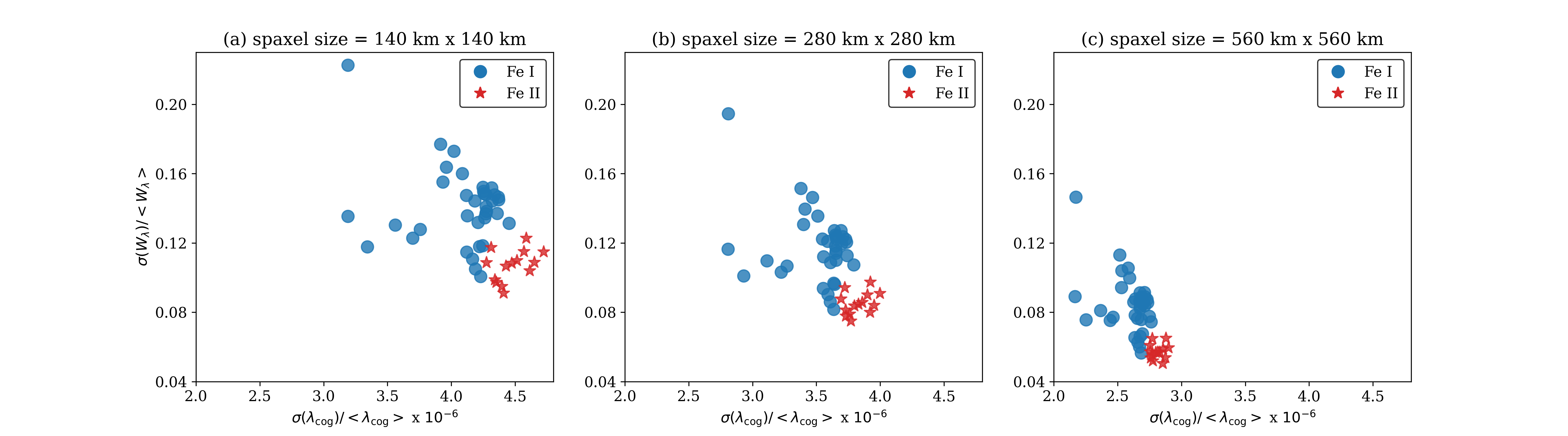}
    \caption{Dependence of the spectral line sensitivity on the size of the spaxel for the line sample given in Tables~\ref{tab:datafei}~and~\ref{tab:datafeii}. The variability in line equivalent width is plotted as a function of the variability in center-of-gravity wavelength displacements similar to \fref{fig:sensitivity}. Sensitivity obtained when 8\,x\,8 pixels (panel a), 16\,x\,16 pixels (panel b) and 32\,x\,32 pixels (panel c) are combined to form a spaxel. The area covered by spaxel in each scenario is indicated in the titles to the panels.}
    \label{fig:spaxelsize}
\end{figure}
\begin{figure}
    \centering
    \includegraphics[scale=0.38]{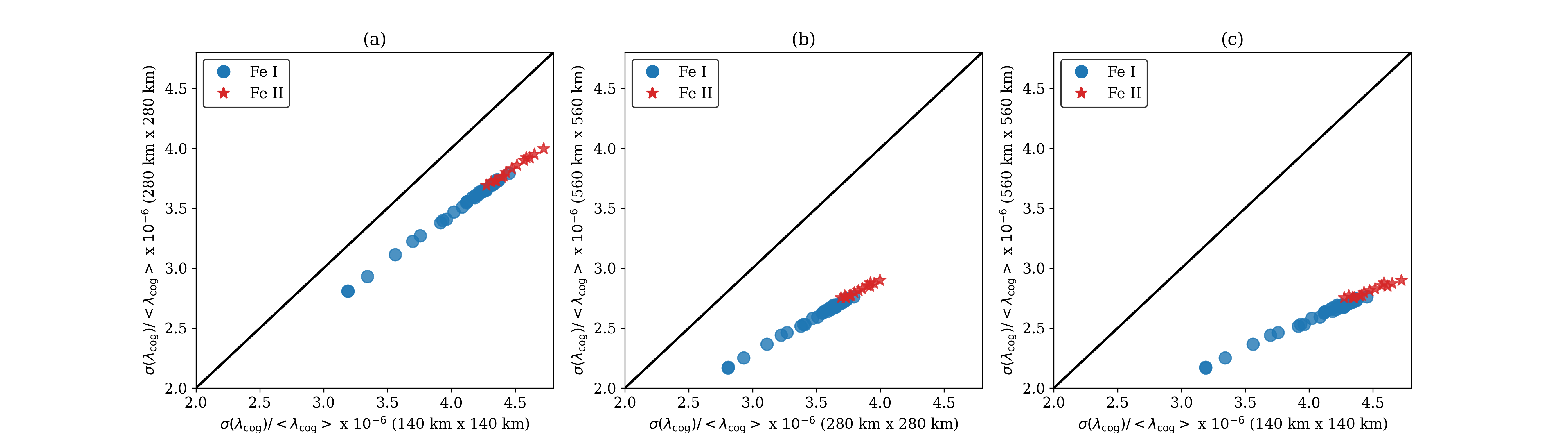}
    \includegraphics[scale=0.38]{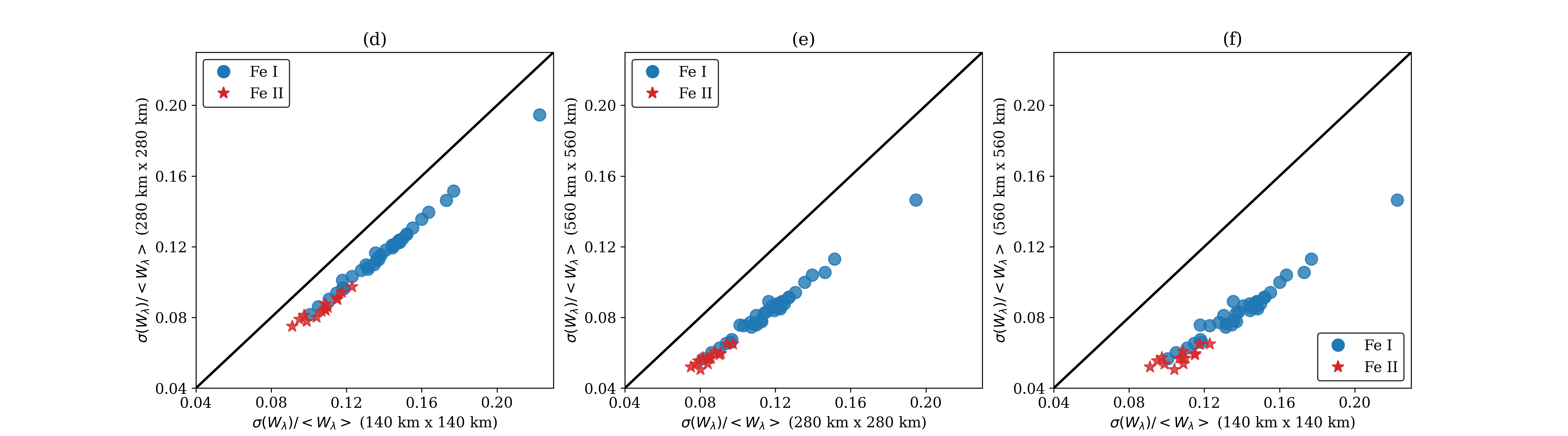}
    \caption{Dependence of the spectral line sensitivity on the size of the spaxel for the line sample given in Tables~\ref{tab:datafei}~and~\ref{tab:datafeii}. Top row: scatter plots of variability in center-of-gravity position for different sizes of spaxels as indicated in axes titles. bottom row: scatter plots of variability in equivalent width for different sizes of spaxels indicated in the axes titles. Circles represent neutral Fe lines while stars represent singly ionized Fe lines.}
    \label{fig:spaxelsize_scatter}
\end{figure}

\section{Single snapshot vs. time series}
\label{app:singlesnap}
\setcounter{figure}{0}
In \fref{fig:timevariation}, we present the response of Fe\,I and Fe\,II to granulation for 45 MURaM snapshots which represent the evolution of solar granulation during a period of 1 hour. The snapshot-to-snapshot variations in the equivalent width and center-of-gravity wavelength displacement variability as plotted as $1\sigma$ error bars. The relatively small error bars, which further reduce by a factor of 6.7 when looking at standard error of the mean, once again confirms that a single snapshot resolving a few dozen granules is sufficient to characterize the variability expected from granulation. 

\begin{figure}
    \centering
    \includegraphics[scale=0.5]{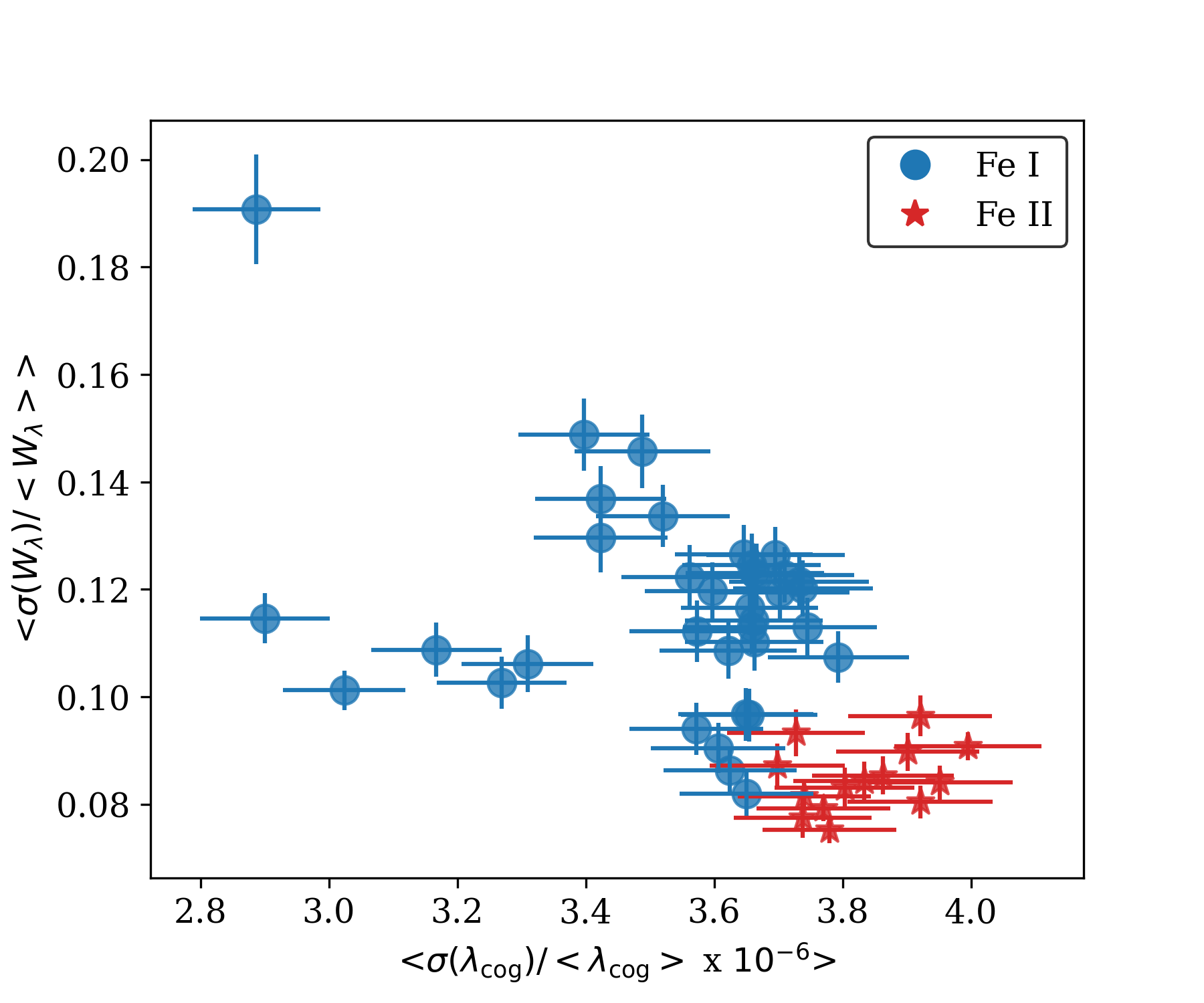}
    \caption{Time variation of the line sensitivity to granulation. Circles represent neutral Fe lines while stars represent singly ionized Fe lines. The error bars represent the standard deviation in equivalent width and center-of-gravity sensitivities calculated for 45 MURaM snapshots covering 1 hour of solar time. The errors shown here reduce by a factor of $\sqrt{45} \approx 6.71$ when the standard error of the mean is calculated, indicating that a single snapshot is sufficient to capture the variability of the granulation pattern.}
    \label{fig:timevariation}
\end{figure}

\section{Sensitivity away from the disk center}
\label{app:viewingangle}
\setcounter{figure}{0}
The spectral line sensitivities at different $\mu$ relative to the ones at $\mu=1.0$ are illustrated in Figures~\ref{fig:clv-cog}~and~\ref{fig:clv-eqw} for the line samples in Tables~\ref{tab:datafei} and \ref{tab:datafeii}. The amplitude of center-of-gravity variations initially increases away from the disk center followed by a decline towards the limb (see \fref{fig:clv-cog}). The amplitude of equivalent width variations increases towards the limb (see \fref{fig:clv-eqw}). 

While the line sensitivities show a moderate dependence on $\mu$, the relative sensitivities of Fe\,I and Fe\,II lines are similar to those at the disk center, i.e. down to $\mu=0.6$ the clouds of points in \fref{fig:clv-cog} do not show much of a spread in the direction perpendicular to the 1:1 line. Thus, we expect that the analysis of the relative line sensitivities presented in our study is representative of the disk-integrated case. We note that the contribution from near-limb regions is in any case rather small: for example, the part of the disk corresponding to  $\mu <0.6$ has a projected area smaller than 20\% of the full disk projected area, and its contribution to the disk-integrated spectrum is further reduced by limb darkening. 

A rather sophisticated center-to-limb variations of the line sensitivities to granulation can be attributed to a multitude of effects caused by the line formation region shifting to higher photospheric layers with decrease of the $\mu$ value and change of the viewing geometry. First, the sensitivity of Fe\,I and Fe\,II concentration to temperature increases towards higher, less dense and cooler layers (property of the Saha equation). This leads to a bigger difference between the granular and intergranular line components and, correspondingly, increase of amplitude of equivalent width variations. Second, temperature contrast between granules and integranules drops towards higher layers. This leads to a more equal weighting (given by continuum intensity $I_c$, see Equations~\ref{eq:cogscatter} and \ref{eq:ewscatter}) of granules and intergranules to the resulting line profile. Third, the difference between mean upward velocity in granules and downward velocity in intergranules also drops with height leading to a smaller shift between the granular and intergranular line components and, thus, decreasing the amplitude of the  center-of-gravity position variations. Finally, horizontal convective flows (that are in general even stronger than vertical; see e.g. \citealt{BellotRubio2009,Obaetal2020}) start to contribute to line-of-sight velocity away from the disk center, increasing the amplitude of the center-of-gravity position variations. This effect acts in the direction opposite to the height-dependence of the vertical velocity. The interplay of these competing effects makes it difficult to assess their relative importance based on qualitative arguments alone, although their combined impact is naturally captured by our calculations shown in Figures~\ref{fig:clv-cog}~and~\ref{fig:clv-eqw}.

\begin{figure}
    \centering
    \includegraphics[scale=0.38]{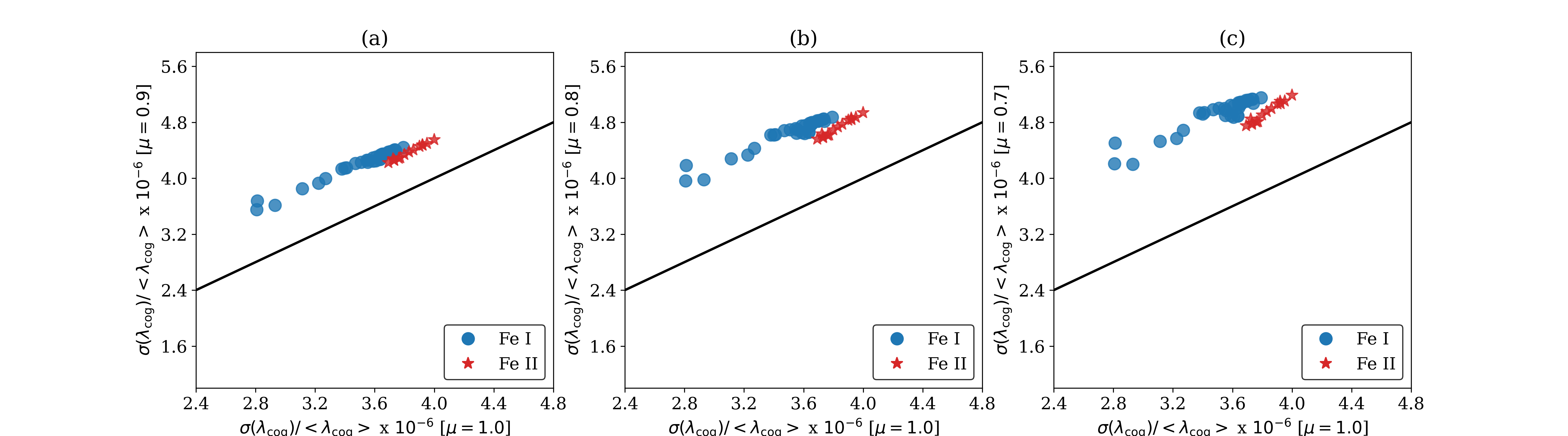}
    \includegraphics[scale=0.38]{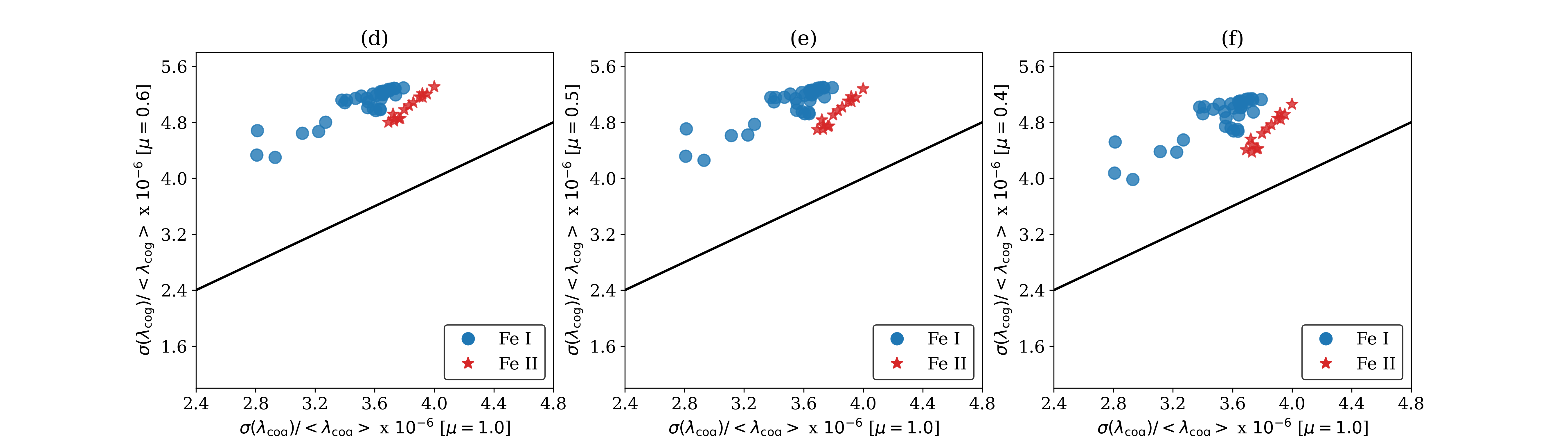}
    \includegraphics[scale=0.38]{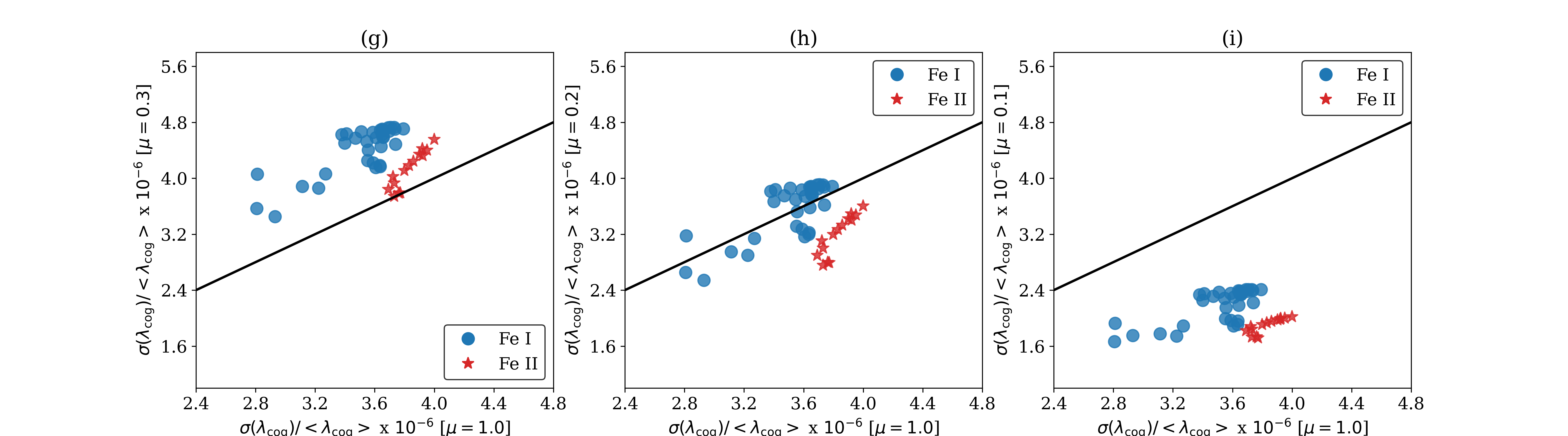}
    \caption{$\mu$-dependence of the line sensitivity to granulation. Panels a-i: variability in center-of-gravity positions for different $\mu$ values as indicated in y-axes titles are plotted against the variability at $\mu=1.0$. Circles represent neutral Fe lines while stars represent singly ionized Fe lines. Solid black line indicates 1:1 line. }
    \label{fig:clv-cog}
\end{figure}

\begin{figure}
    \centering
    \includegraphics[scale=0.38]{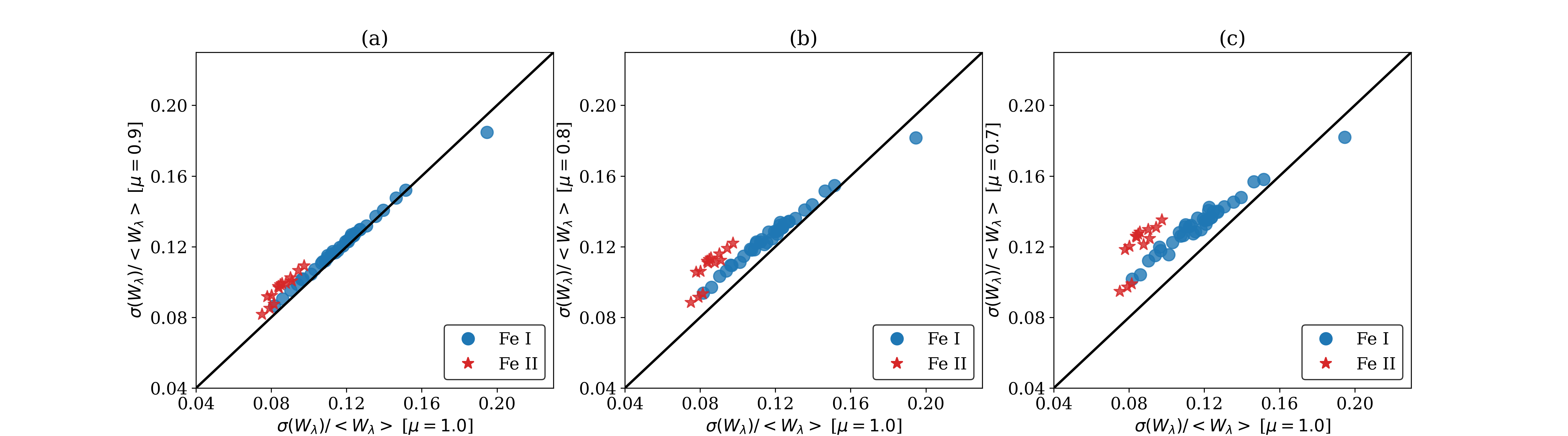}
    \includegraphics[scale=0.38]{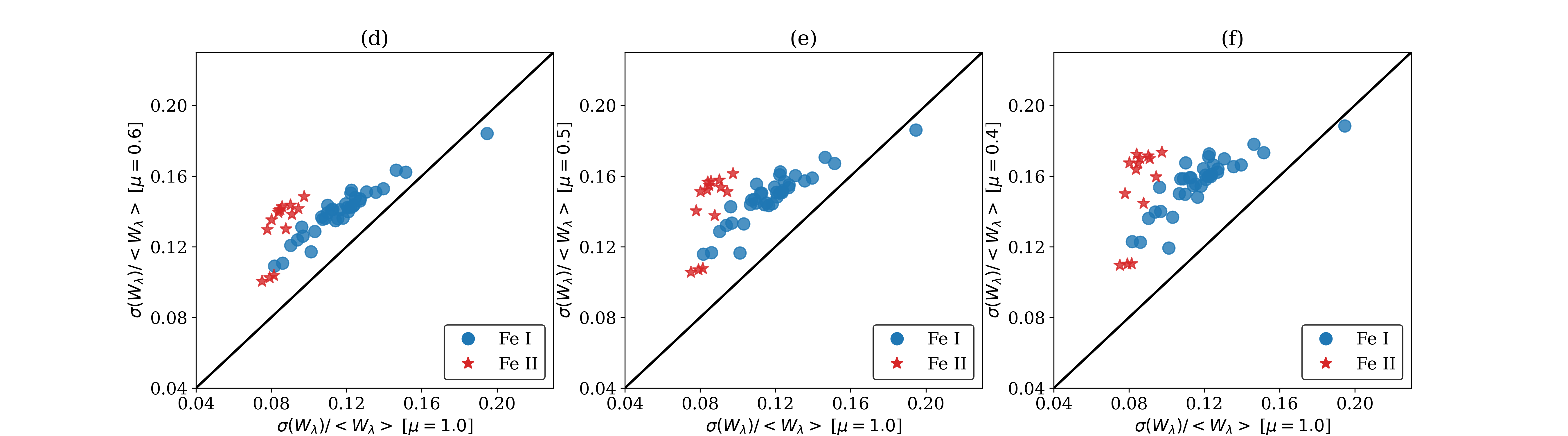}
    \includegraphics[scale=0.38]{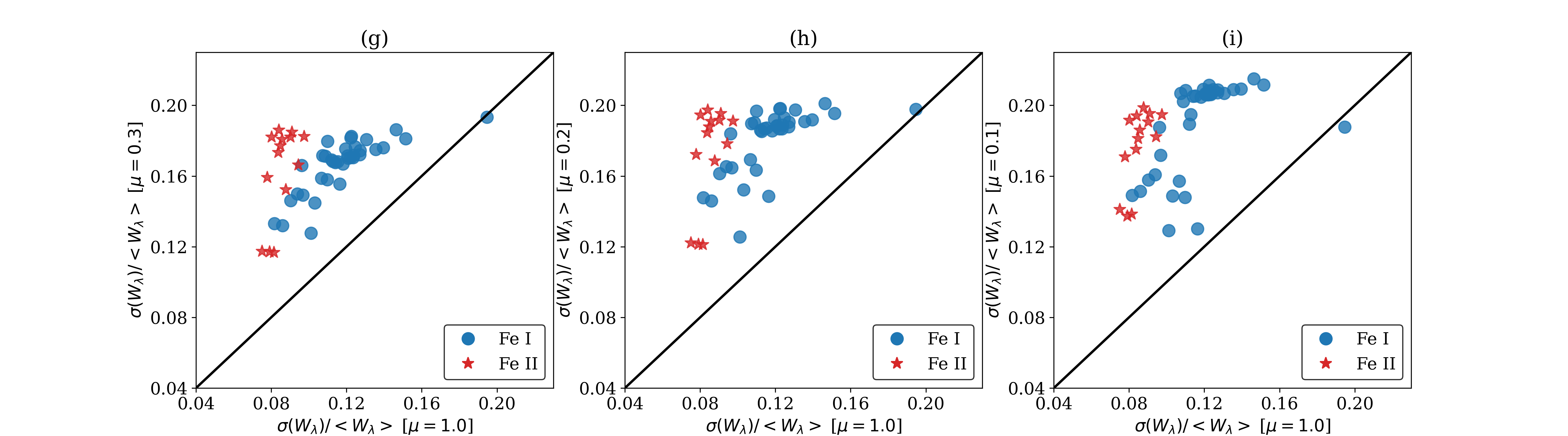}
    \caption{$\mu$-dependence of the line sensitivity to granulation. Panels a-i: variability in equivalent width for different $\mu$ values as indicated in y-axes titles are plotted against the variability at $\mu=1.0$. Circles represent neutral Fe lines while stars represent singly ionized Fe lines. Solid black line indicates 1:1 line. }
    \label{fig:clv-eqw}
\end{figure}

\section{Wavelength dependence of the line sensitivity to granulation}
\label{app:wavelength}
\setcounter{figure}{0}
To investigate how line sensitivity to granulation depends on wavelength, we calculated the sensitivity metrics for disk center using the same sowmium sample discussed in \sref{sec:results-sowmium}, but by changing the line center wavelength from 629.9532\,nm to 400.6515\,nm (blueward) and 800.0445\,nm (redward). We find that for both neutral and singly ionized sowmium lines, the amplitude of equivalent width variations increase at longer wavelengths (see \fref{fig:sowmium-wave-scatter}a,c). The amplitude of center-of-gravity position variations, however, decrease with increasing wavelength (see \fref{fig:sowmium-wave-scatter}b,d).

Here, due to the increase of the continuum opacity due to negative hydrogen ion with wavelength, continuum formation (and to some extent line formation) regions shift to higher layers for longer wavelengths. As a result of this, the line sensitivities are governed by the same effects described in \app{app:viewingangle}, with two exceptions. First, the horizontal velocities no longer matter. Second, the granular and intergranular line components are weighted more equally for longer wavelengths due to the drop of Planck function sensitivity to temperature.

\begin{figure}
    \centering
    \includegraphics[scale=0.5]{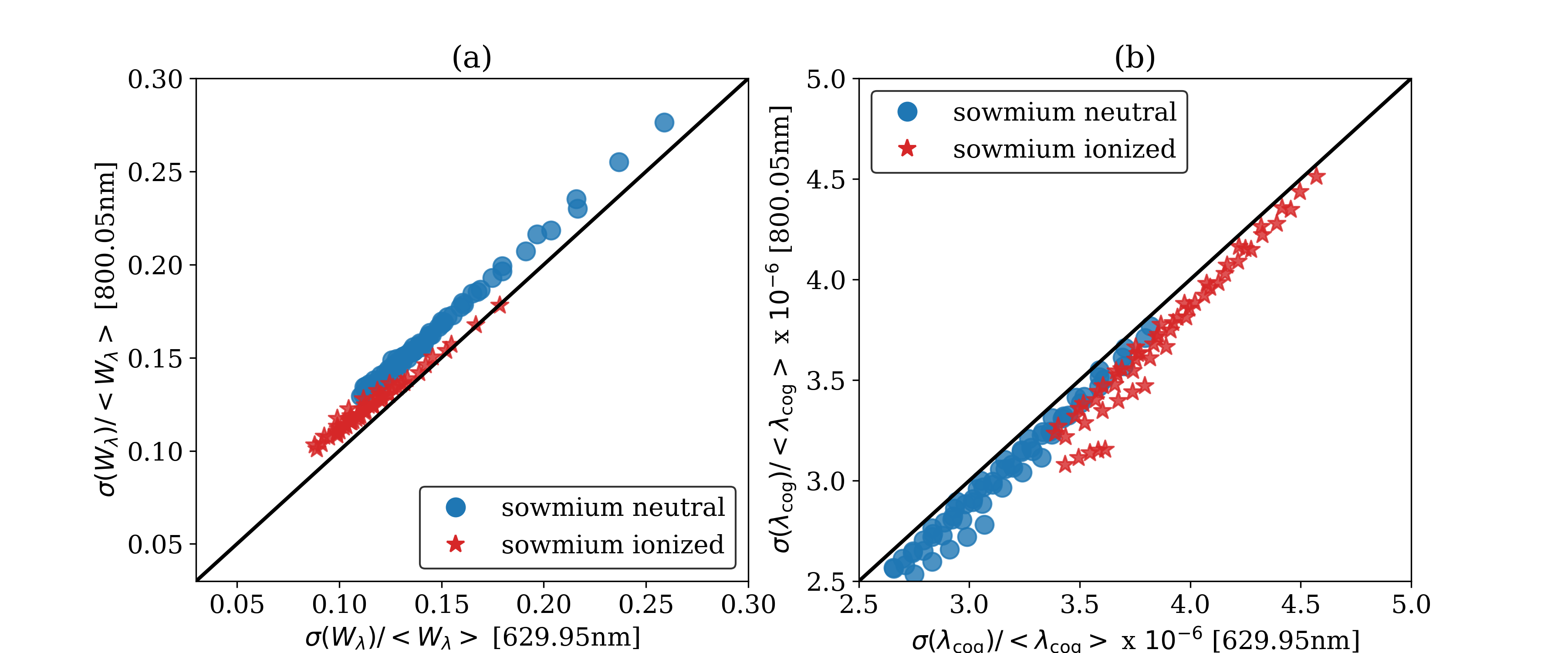}
    \includegraphics[scale=0.5]{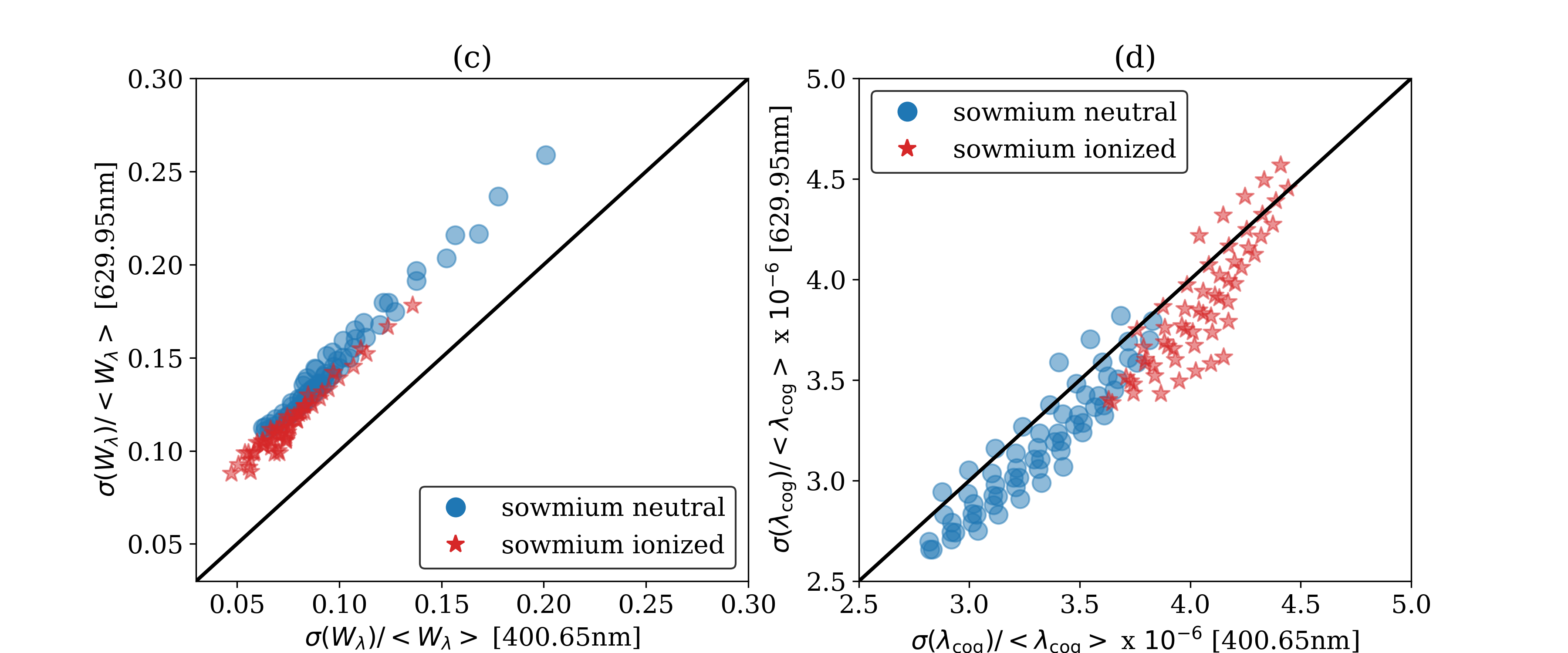}
    \caption{Wavelength dependence of the sowmium line sensitivities at disk center. Panels (a and c): line equivalent width sensitivity at wavelengths indicated in the axes labels. Panels (b and d): line center-of-gravity position sensitivity at the indicated wavelengths. Circles represent neutral sowmium lines while stars represent singly ionized sowmium lines. Solid black line indicates 1:1 line.}
    \label{fig:sowmium-wave-scatter}
\end{figure}

\bibliography{solar}
\end{document}